\documentclass[12pt]{article}
\usepackage[margin=1.25in]{geometry}

\usepackage{natbib}
\usepackage{amsmath, bm}
\usepackage{amssymb}
\usepackage{graphicx}
\usepackage{multicol}
\usepackage{float}
\usepackage{hyperref}
\usepackage{color}
\usepackage{rotating}
\usepackage{subfig}
\usepackage{enumitem}
\usepackage{mathtools}
\usepackage{setspace}
\usepackage{verbatim}
\RequirePackage{etoolbox}
\RequirePackage{xcolor}
\definecolor{cb-blue}{RGB}{0, 109, 219}
\definecolor{cb-rose}{RGB}{255, 109, 182}

\DeclarePairedDelimiter\abs{\lvert}{\rvert}%
\makeatletter
\let\oldabs\abs
\def\abs{\@ifstar{\oldabs}{\oldabs*}}

\makeatletter
\def\namedlabel#1#2{\begingroup
    #2%
    \def\@currentlabel{#2}%
    \phantomsection\label{#1}\endgroup
}
\makeatother

\begin{document}

\title{\large \bf Bayesian Data Synthesis and Disclosure Risk Quantification: An Application to the Consumer Expenditure Surveys}

\author{Jingchen Hu\footnote{Vassar College, Box 27, 124 Raymond Ave, Poughkeepsie, NY 12604, jihu@vassar.edu} $\,$ and Terrance D. Savitsky\footnote{U.S. Bureau of Labor Statistics, Office of Survey Methods Research, Suite 5930, 2 Massachusetts Ave NE Washington, DC 20212, Savitsky.Terrance@bls.gov}}

\maketitle

\begin{abstract}
The release of synthetic data generated from a model estimated on the data helps statistical agencies disseminate respondent-level data with high utility and privacy protection. Motivated by the challenge of disseminating sensitive variables containing geographic information in the Consumer Expenditure Surveys (CE) at the U.S. Bureau of Labor Statistics, we propose two non-parametric Bayesian models as data synthesizers for the county identifier of each data record: a Bayesian latent class model and a Bayesian areal model. Both data synthesizers use Dirichlet Process priors to cluster observations of similar characteristics and allow borrowing information across observations. We develop innovative disclosure risks measures to quantify \emph{inherent} risks in the confidential CE data and how those data risks are ameliorated by our proposed synthesizers. By creating a lower bound and an upper bound of disclosure risks under a minimum and a maximum disclosure risks scenarios respectively, our proposed inherent risks measures provide a range of acceptable disclosure risks for evaluating risk level in the synthetic datasets.
\end{abstract}


\section{Introduction}
\label{intro}
The U.S. Bureau of Labor Statistics (BLS) utilizes various survey programs to collect individual-level and business establishment-level data.  The Consumer Expenditure Surveys (CE) at the BLS is a survey program that focuses on collecting and publishing information about expenditures, income, and characteristics of consumers in the United States. The CE publishes summary, domain-level statistics used for both policy-making and research, including the most widely used measure of inflation - the Consumer Price Index (CPI), measures of poverty that determine thresholds for the U.S. Governments Supplemental Poverty Measure, estimation of the cost of raising a child for making policies on foster care and child support, and estimation of American spending on health care, to name a few.

The CE consists of two surveys: i) the Quarterly Interview Survey, which aims to capture large purchases (such as rent, utilities, and vehicles), contains approximately 7000 interviews, and is conducted every quarter; and ii) the Diary Survey, administered on an annual basis, focuses on capturing small purchases (such as food, beverages, and tobacco), contains approximately 14,000 interviews of households.  In this project, we focus on the CE public-use microdata (PUMD) that result from these instruments. Unlike published CE tables, which release information from the CE data in aggregated forms, the CE PUMD releases the CE data at the individual, respondent level, which potentially enables the CE data users to conduct research tailored to their interests.  Directly releasing individual-level data, however, poses privacy concerns. Under the U.S. Title 13, released versions of public-use data are subject to privacy and confidentiality protection.  Values for sensitive variables deemed at high risk for privacy protection are often suppressed and not reported.

A class of approaches for encoding privacy protection into sensitive variables to permit their public release constructs a Bayesian model for the respondent-level variable(s), estimated on the confidential data, from which new data are simulated or ``synthesized" from the estimated model. The new data, commonly called ``synthetic data", are then proposed for release to the public. The synthetic data generated from flexible models, called synthesizers, should maintain a high level of usefulness (commonly called utility), while smoothing of the data distribution induced by the model often achieves a high level of privacy and confidentiality protection. \citet{Drechsler2011book} provides a comprehensive review of synthetic datasets for statistical disclosure control.

The current CE PUMD of the Interview Survey contains variables about characteristics of the consumer units (CU, i.e. households) and CU members, and detailed tax, income, and expenditure information about the CUs and their members. While rich and useful, a set of important variables about the detailed location of the CUs is not currently released due to privacy concerns and other considerations.  In this paper, we construct synthesizers for the county labels variable for CUs, along with new disclosure risk measures to ensure adequate privacy protection for synthetically-generated county labels, while at the same time ensuring that the synthetic data are useful to the CE data users for various research purposes of interest to them.  Tailored for categorical variables present in the CE data, we propose two non-parametric Bayesian models as data synthesizers. The first synthesizer employs a Dirichlet Process mixtures of products of multinomials (DPMPM), which directly models the county labels variable as a categorical variable where each county in the data receives a unique code. The second synthesizer constructs a new, nonparametric version of areal models with Dirichlet Process priors (DP-areal), which models \emph{counts} of county labels of observations sharing similar characteristics, such as gender, income, and age categories. We choose non-parametric models because they are flexible. Such flexibility is desired because we cannot anticipate what uses (i.e. analyses) that the data analysts will have.

On utility of the synthetic datasets, we demonstrate and compare the effectiveness of the proposed DPMPM and DP-areal synthesizers in preserving important global and local distributional characteristics of the CE data. On disclosure risks evaluation (i.e. evaluating the risks of disclosure by releasing the synthetic data, as the level of disclosure risks is negatively proportional to privacy protection), we propose new disclosure risks measures to capture inherent risk in the original, confidential data to help set context for the reduction in disclosure risks offered by our two candidate synthesizers. Specifically, we consider the inherent minimum and maximum disclosure risks, which give us a lower bound and an upper bound of acceptable disclosure risks. 

\subsection{The CE data}
\label{intro:CEdata}

The CE data sample in our application comes from the 2017 1st quarter Interview Survey. There are $n = 6208$ consumer units (CU) in this sample. We focus on synthesizing the sensitive county variable.

To include predictors, since most variables in the internal CE data suffers from high rate of missingness, Economists on the team that manages the CE survey have suggested a subset of fully-observed variables that are historically good at predicting county labels: gender, income, and age.  After some experimentation, we validated their predictive utility.

Gender is categorical, with 2 levels as coded in the CE data sample. Income and age are discretized, with 4 levels and 5 levels respectively, based on the recommendations by the economists on the CE program. These three variables are non-geographic variables. See Table \ref{tab1} for details of the variables. Working with this set of variables allows the effective use of the DP-areal synthesizer, which limits itself to a moderate set of multi-level predictors due to its computational scalability, and that we show in the sequel produces a robust risk-utility trade-off.

\begin{table}
\caption{Variables in the CE data sample.\label{tab1}}
\begin{tabular}{l l }
\hline
Variable &  Description  \\ \hline
Gender &  Gender of the reference person; 2 categories.\\
Income & Imputed and reported income of the CU; 4 categories (based on 4 quartiles).\\
Age & Age of the reference person; 5 categories ($<$20, 20-40, 40-60, 60-80, $>$80). \\
County & County label of the CU; 133 categories. \\ \hline
\end{tabular}
\end{table}

The county variable represents the county labels of the CUs. As a variable containing detailed geographic information, it is currently \emph{not} released in the CE PUMD for privacy protection. In the 2017 1st quarter CE data sample, there are 133 counties observed, i.e. 133 counties are sampled. These 133 counties are only a small subset of the total number of counties and county-equivalents in the US (3,142 counties and county-equivalents in 2018). The observed 133 counties are scattered around across the nation.  Their sparsity in geographic locations results in the county labels carrying little geographic information. Therefore, we consider county as a categorical variable with 133 levels. We define a ``pattern" as a unique composition of non-geographic variables, i.e., a pattern is determined by intersection of categories for the three non-geographic variables \{Gender, Income, Age\}. The cross tabulation of these three non-geographic variables creates 40 different patterns in total ($2\times4\times5=40$).

As evident in Table 1 of the list of 40 patterns in the Supplementary Material, the number of observations in every pattern varies greatly, from the maximum 454 observations in Pattern 18 to the minimum 3 observations in Patterns 6, 11, and 36.  The presence of patterns with a small number of observations motivates us to develop data synthesizers that allow borrowing information across patterns to strengthen estimation for patterns with small numbers of observations.

\subsection{Literature review}
\label{intro:literature}

\subsubsection{Synthetic data}
\label{intro:literature:synthetic}
Depending on the sensitivity levels of the variables in a study, statistical agencies can either generate fully synthetic datasets, where all variables are deemed sensitive, therefore synthesized \citep{Rubin1993synthetic}, or generate partially synthetic datasets, where only some variables are deemed sensitive and synthesized, and other variables are un-synthesized \citep{Little1993synthetic}. In the CE data sample, since only the county label is deemed sensitive, we aim to generate partially synthetic data where only the county is synthesized. Gender, income, and age are not synthesized.  The record label is maintained in partially synthetic data, though the synthesized variable is generated from the posterior predictive distribution of the synthesizer.

\subsubsection{Synthesis of labels of locations}
\label{intro:literature:locations}
In general, variables containing geographic information are deemed sensitive; however, these variables are extremely helpful for data analysts to conduct research related to locations. Therefore, a number of researchers have proposed synthesizers to generate synthetic geographic data.

One stream of work has treated the geographic location as variable(s) carrying little geographic information, therefore their proposed synthesizers do not incorporate spatial modeling. \citet{WangReiter2012, DrechslerHu2018} developed CART models \citep{Reiter2005CART} to synthesize continuous longitude and latitude. In addition, \citet{DrechslerHu2018} combined the continuous longitude and latitude variables into a single categorical geographic variable, and used versions of categorical CART models for its synthesis. The authors used the DPMPM synthesizer \citep{HuReiterWang2014PSD} to generate synthetic categorical locations.

Another stream of work has explicitly incorporated spatial modeling in their synthesizers for densely-observed geographic variables. \citet{Paiva2014SM} aggregated counts of geographic locations to a pre-defined grid level, modeled the counts through areal level spatial model, which included spatial random effects with Conditional Autoregressive (CAR) priors \citep{Besag1991car}, then synthesized counts of locations from the estimated model to release. \citet{QuickHolanWikleReiter2015SS} developed synthesizers based on Bayesian marked point process \citep{LiangCarlinGelfand2009}. \citet{ZhouDominiciLouis2010, QuickHolanWikle2018JRSSA} developed differentially smoothing-based synthesizers based on spatially-indexed distances.

Considering the county labels in the CE data sample, especially the fact that the observed 133 county labels is only a small subset (a little over 4\%) of the total number of 3,142 counties and county-equivalents in the US, we believe the county labels themselves carry little spatial correlation information. Therefore, we work with synthesizers that do not incorporate spatial modeling. Specifically, we choose the DPMPM synthesizer, and we develop a new, non-spatial version of the count-based areal synthesizer in \citet{Paiva2014SM} with non-spatial, nonparametric priors, labeled as DP-areal synthesizer.

\subsubsection{Disclosure risks}
\label{intro:literature:risks}
Developing appropriate synthesizers allows statistical agencies to generate useful synthetic datasets; however, before the release of synthetic data to the public, the data disseminating agencies have to evaluate the level of privacy protection (or the lack of) provided by the synthetic data. Synthetic data release takes place only if the synthetic data expresses a sufficient level of privacy protection. Typically, privacy protection in the synthetic data is determined by the evaluation of its disclosure risks. The higher the disclosure risks, the lower the privacy protection, and vice versa.

There are two general categories of disclosure risks in synthetic datasets: i) identification disclosure risks, and ii) attribute disclosure risks. For partially synthetic data, such as only synthesizing the county labels and keeping other non-geographic variables un-synthesized in the CE data sample, both identification disclosure risks and attribute disclosure risks exist \citep{Hu2018TDP}.

Identification disclosure risks exist when an intruder has access to information about some of the variables for a target record through external files, and tries to match those values with available information in the released data to identify the name associated to that record. Widely used measures for this type of risks are based on Bayesian probabilistic matching \citep{DuncanLambert1986, DuncanLambert1989, Lambert1993, Reiter2005risk, ReiterMitra2009}; for example, in the CE data sample, suppose an intruder knows the particular age and gender combination, as well as the county label, of a person named ``Betty".  The intruder wants to identify which record in the CE synthetic data belongs to Betty. Intuitively, the intruder will go through the synthetic datasets by matching the known age and gender combination, and the synthesized county label. Suppose record, $i$, belongs to Betty.  Let $c_i$ denote the number of records in the sample sharing the same combination and the \emph{true} county label (in the original data) for record $i$. Then $1/c_i$ gives a probability of the intruder randomly and correctly guessing the record attached to Betty based on matching attribute values. In general, the larger the value of $c_i$, the lower the identification disclosure risks for record $i$.  If the county label in the released synthetic data for record $i$ is different from the true label in the real data, then record $i$ will not be among the $c_i$ records, which means Betty is not among the those records.  Let $T_{i} \in \{0,1\}$ be a binary indicator of whether the true match is among the $c_{i}$ records.   If the county label in the synthetic data for Betty's record, $i$, is the same as the real data, then $T_{i}=1$; otherwise, if the county label in the synthetic data for record $i$ is different from that in the real data, then $T_{i} = 0$.  If $T_{i} = 1$, the intruder has a $1/c_{i}$ probability of finding the record belonging to Betty.  If $T_{i} = 0$, however, the intruder has a $0$ probability of finding the record belonging to Betty because her record is not among the $c_{i}$.   Therefore, the ratio $T_{i}/c_{i}$ provides a measure of expected identification match risk for record $i$.

In addition to the expected match risks, measures such as the true match rate (the percentage of true unique matches among target records) and the false match rate (the percentage of false matches among unique matches) are also useful \citep{ReiterMitra2009, DrechslerReiter2010, HuHoshino2018PSD, Hu2018TDP, DrechslerHu2018}.

Attribute disclosure risk measures how likely an intruder is to discover the true value of synthesized variables; in our case the county label. \citet{Reiter2012discussion, WangReiter2012, Reiter2014framework} proposed a Bayesian approach to compute a posterior probability of identifying the true attribute for each record under the synthesizer \citep{HuReiterWang2014PSD, Paiva2014SM, WeiReiter2016, HuReiterWang2018BA, ManriqueHu2018JRSSA}. This general framework has the advantages of providing interpretable probability statements of the attribute disclosure risks \citep{Hu2018TDP}. Yet, the procedure requires that the intruder knows the true attribute values of the synthesized variables for every other record except the target record in order to make the approach computationally tractable.  The approach does not scale up well to multiple categorical synthesized variables or a synthesized variables with many categories, such as our county label variable (with $133$ categories).  In our synthetic CE data application, we construct an attribute disclosure risk measure from our identification risk formulation.  Our approach counts the number of records in the file where the synthesized value matches the true data value to provide an overall file level summary that contrasts with the record level statistic in \citet{HuReiterWang2014PSD}.

Moreover, in the sequel we extend these identification and attribute disclosure risk measures in a novel way to capture disclosure risks inherent in the real data, independent of the synthesizers. We create risk measures under a minimum scenario and a maximum scenario, and obtain a lower bound and an upper bound, respectively.  We argue that disclosure risks in the synthetic datasets may be judged based on how they fit within these bounds.

The remainder of the paper is organized as follows: In Section \ref{synthesizers}, we describe the DPMPM synthesizer, the DP-areal synthesizer, and the computation details of their implementation. Section \ref{utility} presents the utility measures and results of the synthetic CE; Section \ref{risks} presents the proposed disclosure risks measures for the original CE data, and the disclosure risks results of the synthetic CE data. The paper concludes with discussion in Section \ref{conclusion}.

\section{Synthesizers}
\label{synthesizers}

Our proposed synthesizers focus on generating partially synthetic data for multivariate categorical data. The synthesizers will allow non-zero probabilities for unobserved combinations of variables, provided the unobserved combinations are not structural zeros (i.e. impossible combinations). We proceed to describe two synthesizers in general settings, with illustrations of synthesizing the country label attribute in the CE data sample.


\subsection{Dirichlet Process mixtures of product of multinomials (DPMPM)}
\label{synthesizers:DPMPM}

The DPMPM is a Bayesian version of latent class models for unordered categorical data. The Dirichlet Process prior specifies infinite number of mixtures, allows the data to learn the effective number of mixture components, and provides support to all distributions of multivariate categorical variables \citep{DunsonXing2009JASA}. \citet{SiReiter2013JEBS} used the DPMPM as a missing data imputation engine, and \citet{HuReiterWang2014PSD} first used it as a synthesizer for a sample of American Community Survey (ACS). \citet{DrechslerHu2018} also used the DPMPM synthesizer for simulating geolocations of a large scale administrative data in Germany.

The description of the DPMPM synthesizer follows that in \citet{HuHoshino2018PSD}. Consider a sample $\mathbf{X}$, that consists of $n$ records, and each record has $p$ categorical variables. For the CE data, $p = 4$, including non-geographic variables (gender, income, and age), and geographic variable, the county label. The basic assumption of the DPMPM is that every record, $\mathbf{X}_i = (X_{i1}, \cdots, X_{ip}),~i \in (1,\ldots,n)$, belongs to one of $K$ underlying latent classes, which is unobserved, thus latent. Given the latent class assignment $z_i \in (1,\ldots,K)$ of record $i$, as in Equation (\ref{eq:DPMPM2}), the value for record $i$ and attribute, $j \in (1,\ldots,p)$, $X_{ij}$, independently follows its own multinomial distribution, as in Equation (\ref{eq:DPMPM1}), where $d_j$ denotes the number of categories of variable $j$.
\begin{eqnarray}
	\label{eq:DPMPM1} X_{ij} \mid z_i, \theta &\overset{ind}{\sim}& \textrm{Multinomial}(1; \theta_{z_i1}^{(j)}, \dots,
		\theta_{z_i d_j}^{(j)}) \,\,\,\, \textrm{for all } i, j,\\
	\label{eq:DPMPM2} z_i \mid \pi &\sim& \textrm{Multinomial}(1; \pi_1, \dots, \pi_K) \,\,\,\, \textrm{for all } i.
\end{eqnarray}

The DPMPM clusters records with similar characteristics based on \emph{all} $p$ attributes. Relationships among these $p$ categorical attributes are induced by integrating out the latent class assignment $z_i$. To empower the DPMPM to pick the effective number of occupied latent classes, the truncated stick-breaking representation \citep{Sethuraman1994SS} is  used as in Equation (\ref{eq:DPMPMprior-pi}) through Equation (\ref{eq:DPMPMprior-theta}),
\begin{eqnarray}
\label{eq:DPMPMmodelprior}
	\label{eq:DPMPMprior-pi} \pi_k &=& \beta_{k}\prod_{l<k}(1-\beta_{l}) \,\,\,\, \textrm{for } k=1, \dots, K,\\
	\label{eq:DPMPMprior-V} \beta_{k}&\overset{iid}{\sim}& \textrm{Beta} (1,\alpha) \,\,\,\, \textrm{for } k=1, \dots, K-1, \,\,\,\, \beta_K=1,\\
	\label{eq:DPMPMprior-alpha} \alpha &\sim& \textrm{Gamma}(a_{\alpha}, b_{\alpha}),\\
	\label{eq:DPMPMprior-theta} \mathbf{\theta}_{k}^{(j)}=(\theta_{k1}^{(j)},\dots,\theta_{kd_j}^{(j)})&\sim& \textrm{Dirichlet} (a_{1}^{(j)}, \dots,a_{d_{j}}^{(j)}),
\end{eqnarray}
and a blocked Gibbs sampler is implemented for the Markov chain Monte Carlo sampling procedure \citep{IshwaranJames2001JASA, SiReiter2013JEBS, HuReiterWang2014PSD, DrechslerHu2018}.

To generate synthetic county label of each record using the DPMPM synthesizer, we first generate sample values of $(\pi^{(\ell)}, \alpha^{(\ell)}, \theta^{(c)(\ell)})$ from the posterior distribution , where $\theta^{(c)(\ell)}$ contains the sample values of the county label variable at MCMC iteration $\ell$.  We can generate the vector of latent class assignments $\{z_i^{(\ell)}, i = 1, \cdots, n\}$ through a multinomial draw with the samples of $\pi^{(\ell)}$, as in Equation ({\ref{eq:DPMPM2}). We next generate synthetic county label, $\{X_{ic}^{(\ell)}, i = 1, \cdots, n\}$, through a multinomial draw with samples of $\theta^{(c)(\ell)}$, as in Equation ({\ref{eq:DPMPM1}). Let $\mathbf{Z}^{(\ell)}$ denote a partially synthetic dataset at MCMC iteration $\ell$. Then we repeat the process for $m$ times, creating $m$ independent partially synthetic datasets $\mathbf{Z} = (\mathbf{Z}^{(1)}, \cdots, \mathbf{Z}^{(m)})$.

One can generate multiple synthetic variables using the DPMPM synthesizer by generating each synthetic variable independently at iteration $\ell$. Suppose there are $s$ ($s \leq p$) variables to be synthesized. We first generate sample values of $(\pi^{(\ell)}, \alpha^{(\ell)}, \{\theta^{(r)(\ell)}\} (r = 1, \cdots, s))$, where $\theta^{(r)(\ell)}$ contains the sample values of the $r$-th variable to be synthesized at MCMC iteration $\ell$. After generating the vector of latent class assignments $\{z_i^{(\ell)}, i = 1, \cdots, n\}$ as before, we next generate synthetic $r$-th variable, $\{X_{ir}^{(\ell)}, i = 1, \cdots, n\}$, through a multinomial draw with samples of $\theta^{(r)(\ell)}$, as in Equation ({\ref{eq:DPMPM1}). One repeats the last step for each of the $s$ variables to be synthesized, and creates a partially synthetic dataset $\mathbf{Z}^{(\ell)}$ at MCMC iteration $\ell$.

\subsection{Areal models with Dirichlet Process prior on random effect (DP-areal)}
\label{synthesizers:DPareal}

The DP-areal synthesizer is built upon areal level spatial models, also known as disease mapping models \citep{ClaytonKaldor1987, Besag1991car, ClaytonBernardinelli1992}. \citet{Paiva2014SM} developed extensions of the areal level spatial models as engines to generate simulated locations. Specifically, they i) created pre-defined areal units based on non-geographic variables, ii) aggregated counts of geographic locations to the pre-defined areal, iii) estimated areal level spatial models that predict observed, areal-level counts with spatial random effects using a Conditional Autoregressive (CAR) prior, and iv) simulated new locations for each individual from the estimated models. Crucial to the setup in \citet{Paiva2014SM} are the pre-defined patterns formed by the intersection of non-arial attributes. Recall that a pattern for the CE sample is determined by the composition of \{Gender, Income, Age\}, and there are 40 patterns in the CE data.

However, as discussed before, due to the little geographic information carried in county labels in the CE data as a result of the geographic sparsity of county labels, the use of spatial random effects and CAR priors on them in \citet{Paiva2014SM} is not appropriate. Instead, we include non-spatial random effects from other sources and specify Dirichlet Process priors for them in our application.  We now turn to the description of our DP-areal synthesizer.

Let $b$ denote a unique pattern of non-geographic variables, and $b = 1, \ldots, B$, where $B$ is the total number of unique patterns ($B = 40$ in the CE data). Let $c_i^{(b)}$ be the count of \emph{observations} in county $i$ \emph{within} pattern $b$. When there is no observation of a particular county $i$ for pattern $b$, zeros are inserted so that $c_i^{(b)} = 0$. For clarity, we reserve the word ``combination" for non-geographic variables {\it{and}} the geographic attribute county label, and the word ``pattern" for only the non-geographic variables. Specifically our model assumes,
\begin{eqnarray}
c_i^{(b)} &\sim& \textrm{Poisson}(\lambda_i^{(b)}), \label{eq:DPareal1}\\
\log \lambda_i^{(b)}  &\sim& \textrm{Normal}(\mu + \theta^*_{z_i^{(b)}} + \sum_{r=1}^R \phi^*_{{z_i^{(b)}}r}X_{i,r}^{(b)}, \frac{1}{\tau_{\lambda}}), \label{eq:DPareal2} \\
z_i^{(b)} \mid \rho &\sim& \textrm{Multinomial}(1; \rho_1, \cdots, \rho_K). \label{eq:DPareal3}
\end{eqnarray}

In the regression $\log \lambda_i^{(b)}  \sim \textrm{Normal}(\mu + \theta^*_{z_i^{(b)}} + \sum_{r=1}^R \phi^*_{{z_i^{(b)}}r}X_{i,r}^{(b)}, \frac{1}{\tau_{\lambda}})$ in Equation (\ref{eq:DPareal2}), $\mu$ is the overall intercept for $\log(\lambda_i^{(b)})$, the logarithm of Poisson rate for county $i$ and pattern $b$. This set-up specifies a Poisson-lognormal model where precision parameter, $\tau_{\lambda}$, allows for over-dispersion.  Note that we let $r = 1, \cdots, R$, where $R = \sum_{j=1}^{p} d_j$, represents the total sum of the number of categories of all non-geographic categorical variables. Subsequently, $\mathbf{X}_{i}^{(b)}$ is an $R \times 1$ vector comprising ones at positions $X_{i,r}^{(b)}$ (the attribute values at positions for all non-geographic attributes in pattern $b$) and zeros elsewhere.

Two types of random effects are considered: i) combination-specific random effect, denoted by $\theta^{\ast}$, and ii) county-specific and variable-specific random effect, denoted by $\phi^{\ast}$. To adequately model these random effects, the truncated DP priors are specified on $\theta^*$'s and $\phi^*$'s to allow flexible clustering counties of similar characteristics.  Here, $z_{i}^{(b)} \in (1,\ldots,K)$, denotes the cluster indicator for each combination, $(i,b)$.  Then, $\theta_{z_i^{(b)}}^{\ast}$ represents the combination-specific random effect, where all combinations in the same mixture component (i.e. when $z_i^{(b)} = z_j^{(b')} = k$) share the same unique random effect value or ``location", $\theta_{k}^{\ast}$. Similarly, $\phi_{z_i^{(b)}r}^{\ast}$ is a county-specific and value-specific random effect, where all counties in the same mixture component (i.e., when $z_i^{(b)} = z_j^{(b')} = k$) share the same random effect, $\phi_{kr}^{\ast}$.  The cluster assignment, $z_{i}^{(b)}$, for each combination is generated from a multinomial draw with cluster probabilities, $\rho_1, \cdots, \rho_K$  in Equation (\ref{eq:DPareal3}) with cluster-indexed coefficients or locations, given by $\left(\theta^{\ast}_{k},(\phi^{\ast}_{kr})_{r=1,\ldots,R}\right)_{k=1,\ldots,K}$.
Moreover, the total number of mixture components or clusters is truncated at $K$. The current truncated mixture model becomes arbitrarily close to a Dirichlet Process mixture as $K \to \infty$ in Equation (\ref{eq:DPareal3}).

We use the truncated stick-breaking representation for the prior distribution of $\rho$ \citep{Sethuraman1994SS}. We specify {\it i.i.d.} normal priors for $\theta^{\ast}$'s and multivariate normal priors for locations, $\phi^{\ast}$'s, and a univariate normal prior for the overall mean $\mu$.
\begin{eqnarray}
\theta^*_k &\overset{iid}{\sim}& \textrm{Normal}(0, 1/\tau_{\theta}), \label{eq:DPareal7} \\
(\phi^*_{k1}, \cdots, \phi^*_{kR}) &\sim& \textrm{MultivariateNormal}_R(\mu_{\phi}, \Sigma_{\phi}), \label{eq:DPareal8} \\
\mu &\sim& \textrm{Normal}(0, 1/\tau_{\mu}). \label{eq:DPareal9}
\end{eqnarray}

To generate the synthetic county label of each record, we follow the general approach of \citet{Paiva2014SM}. At MCMC iteration $\ell$, we, firstly, gather all records with the same pattern. Secondly, we collect all the $\lambda_{i, \ell}^{(b)}$'s from the unique combination of pattern $b$ and the county $i$. Thirdly, we compute
\begin{eqnarray}
p_{i, \ell}^{(b)} = \lambda_{i, \ell}^{(b)} / \sum_{i=1}^{G} \lambda_{i, \ell}^{(b)},
\label{eq:DPareal_syn1}
\end{eqnarray}
where $G$ is the number of all county labels within pattern $b$ (e.g. $G = 133$ in the CE data). Finally, we take a multinomial draw from
\begin{eqnarray}
S_{h, \ell}\sim \textrm{Multinomial}(1; p_{1, \ell}^{(b)}, \cdots, p_{G, \ell}^{(b)}),
\label{eq:DPareal_syn2}
\end{eqnarray}
where $S_{h, \ell}$ is the random variable representing the county label of record $h$, $h \in (1,\ldots, c_i^{(b)})$. We repeat this process for all records in the sample, creating a partially synthetic dataset $\mathbf{Z}^{(\ell)}$. Then the entire process is repeated for $m$ times, creating $m$ independent partially synthetic datasets, $\mathbf{Z} = (\mathbf{Z}^{(1)}, \cdots, \mathbf{Z}^{(m)})$.

Unlike the flexibility of the DPMPM synthesizer to create synthetic values for multiple categorical variables, $s \leq p$ (when $s = p$ it becomes fully synthetic data; see examples in \citet{HuReiterWang2014PSD} for fully synthetic data applications using DPMPM), the DP-areal synthesizer needs to create a combined variable of the $s$ variables to then create counts for modeling. It could face computing and estimation challenges when $s$ is relatively large and each has many categories. 

\subsection{Computation}
\label{synthesizers:computation}

Computation of the DPMPM synthesizer is done by the \texttt{NPBayesImputeCat} R package \citep{NPBayesImputeCat}. We run the DPMPM synthesizer on the CE sample for 10,000 iterations with 5000 burn-in. We follow the recommendations of \citet{DunsonXing2009JASA, SiReiter2013JEBS, HuReiterWang2014PSD, DrechslerHu2018} and set $a_{\alpha} = b_{\alpha} = 0.25$, and set uninformative priors for $\theta$ by $a_1^{(j)} = \cdots = a_{dj}^{(j)} = 1$ for $j = 1, \cdots, p$. We set $K = 40$ and track the number of occupied latent classes with 95\% interval (28, 36), indicating $K = 40$ is sufficiently large. We generate $m = 20$ synthetic datasets by using parameters in iterations that are far away from each other to guarantee independence. We label the $m = 20$ synthetic datasets generated by the DPMPM synthesizer as $\mathbf{Z}_{DPMPM}$. 

Computation of the DP-areal synthesizer is done using Stan programming language \citep{Rstan}. We ran the DP-areal synthesizer on the CE sample for 4000 iterations with 2000 burn-in. Since Stan employs a Hamiltonian Monte Carlo (HMC) sampler that suppresses the usual random walk behavior of the Metropolis-Hastings sampler, posterior sampling iterations are far less correlated than under a Gibbs sampler, which permits use of far fewer iterations.  We set $a_{\alpha} = b_{\alpha} = 1$, and specify $\textrm{Gamma}(1,1)$ prior distributions for $\tau_{\theta}, \tau_{\phi}, \tau_{\mu}$, and $\tau_{\lambda}$. For the multivariate covariance matrix $\Sigma_{\phi}$ in the prior distribution for $\phi^{\ast}$'s, we decompose $\Sigma_{\phi} = (\textrm{I}\bm{\sigma}_{\phi})\Omega_{\phi}(\textrm{I}\bm{\sigma}_{\phi})$, into the  $R\times 1$ vector of variances, $\bm{\sigma}_{\phi}$ that are diagonalized into an $R\times R$ matrix in $\textrm{I}\bm{\sigma}_{\phi})$, and an $R\times R$ correlation matrix, $\Omega_{\phi}$. We select a truncated $\textrm{t}(3,0,1)$ prior distribution for the components of $\bm{\sigma}_{\phi}$.  We choose the $\textrm{LKJ}(\nu = 2)$ prior distribution for $\Omega_{\phi}$, which has a single hyperparameter, $\nu$, that controls how tightly the prior distribution is centered on the identity matrix (meaning independence).  We select $\nu = 2$, which denotes a uniform distribution over the space of correlation matrices, the most weakly informative prior possible, such that we let the data learn the values. We set $K = 50$, and generate $m = 20$ synthetic datasets. They are labeled as $\mathbf{Z}_{DP-areal}$.

Convergence for both model runs is assessed used the Gelman-Rubin test statistic that measures between MCMC chain variance (using 4 chains) versus within chain variance.  A value close to 1 indicates convergence of the MCMC chains.  We confirm convergence and robust mixing by computing the effective sample size (to account for within-chain correlations) and achieve approximately the same effective sample size of 1500 for both the DPMPM synthesizer under Gibbs sampling and the DP-areal synthesizer under HMC.
\section{Utility measure}
\label{utility}


Measures of the level of the closeness between the inference results from the original data and from the synthetic data, commonly referred to as utility measures, are needed to evaluate the usefulness of the synthetic data. We focus on measuring the preservation of distributional characteristics of the synthetic data in the synthetic datasets.

The CE survey program informed us the most common use of the geography information is the table counts by county. Since only the county labels are synthesized in the synthetic CE data, we examine the preservation of the distributional characteristics of the county label, and its relationships with other un-synthesized variables. We do so by conducting the same analysis on the original dataset, and on $\mathbf{Z}_{DPMPM}$ and $\mathbf{Z}_{DP-areal}$ and compare results to the original data, for context.  Since we have defined patterns in the CE data, we will consider three categories of utility measures: i) a globally utility measure, which focuses on the distributional characteristics of county label and its relationships with other variables at the file level; ii) an analysis-specific utility measure, which focuses on inferences the data analysts are likely to make using the released county labels; and iii) a within pattern utility measure, which focuses on the distributional characteristics of county label at the pre-defined pattern level.

\subsection{Global utility measure}

We proceed to formulate measures of utility for our two synthesizers based on a typical manner in which data analysts use the CE data.   In the CE data, three non-geographic variables (gender, income, and age) and one geographic variable (county label) are all constructed to be categorical. Furthermore, only the county label is synthesized. It is therefore useful to calculate one-way, two-way, and three-way tables of counts of observations for the entire file, and compare these computed counts from the original data, to those from the synthetic datasets.  Comparing the accuracy of the synthetic data in reproducing table counts provides a deviance measure of the synthetic datasets from the original dataset. Since these tables are constructed for the entire files consisting all records, this utility measure is regarded as global utility measure. Such measures of utility are generalizable and apply to any categorical data \citep{DrechslerHu2018}.

Without loss of generality, let $\mathbf{Z}$ denote the $m$ synthetic datasets generated from a synthesizer and $\mathbf{X}$ be the original CE data.


For one-way tables, we compute the counts of observations of the 133 categories of county label in $\mathbf{X}$, as well as the counts of observations of the 133 categories of county label in $\mathbf{Z}^{(\ell)},~\ell = 1,\ldots,m$. We next calculate the differences in the counts between the original and synthetic datasets, and report the sum of the absolute differences to avoid cancellation of positive and negative differences. This process is repeated for every $\mathbf{Z}^{(\ell)}$, $\ell = 1, \cdots, m$.
For the two-way tables, we compute the counts of observations in the contingency tables formed by county label and another non-geographic variable and follow the same procedure as for the one-way tables.  Similarly for the three-way tables, counts of observations in the contingency tables formed by county label and two other non-geographic variables are computed in $\mathbf{X}$ and $\mathbf{Z}^{(\ell)}$.  Table \ref{tab:utility_DPMPMandDPareal} gives summary of absolute deviance of the synthetic data from the original data. Results are averages of $m = 20$ partially synthetic datasets, with results on $\mathbf{Z}_{DPMPM}$ in column DPMPM, and results on $\mathbf{Z}_{DP-areal}$ in column DP-areal. These summaries show that the DPMPM synthesizer produces smaller absolute deviance than the DP-areal synthesizer, especially in the one-way and two-way tables.

\begin{table}
\caption{Sum of deviations for each of one-way, two-way, and three-way tables of the synthetic datasets $\mathbf{Z}_{DPMPM}$ and $\mathbf{Z}_{DP-areal}$ from those of the original dataset $\mathbf{X}$. Results are averages of $m = 20$ partially synthetic datasets, divided by 100 for readability.\label{tab:utility_DPMPMandDPareal}}
\centering
\begin{tabular}{r|r|r}
\hline
Table & DPMPM & DP-areal \\
\hline
 one-way & 9.087 & 17.105\\
 two-way &47.908 & 64.165\\
 three-way &80.826 &  88.843\\
\hline
\end{tabular}
\end{table}

\begin{figure}[t]
\centering
    \subfloat[One-way.\label{fig:oneway}]{
        \centering
        \includegraphics[width=0.33\textwidth]{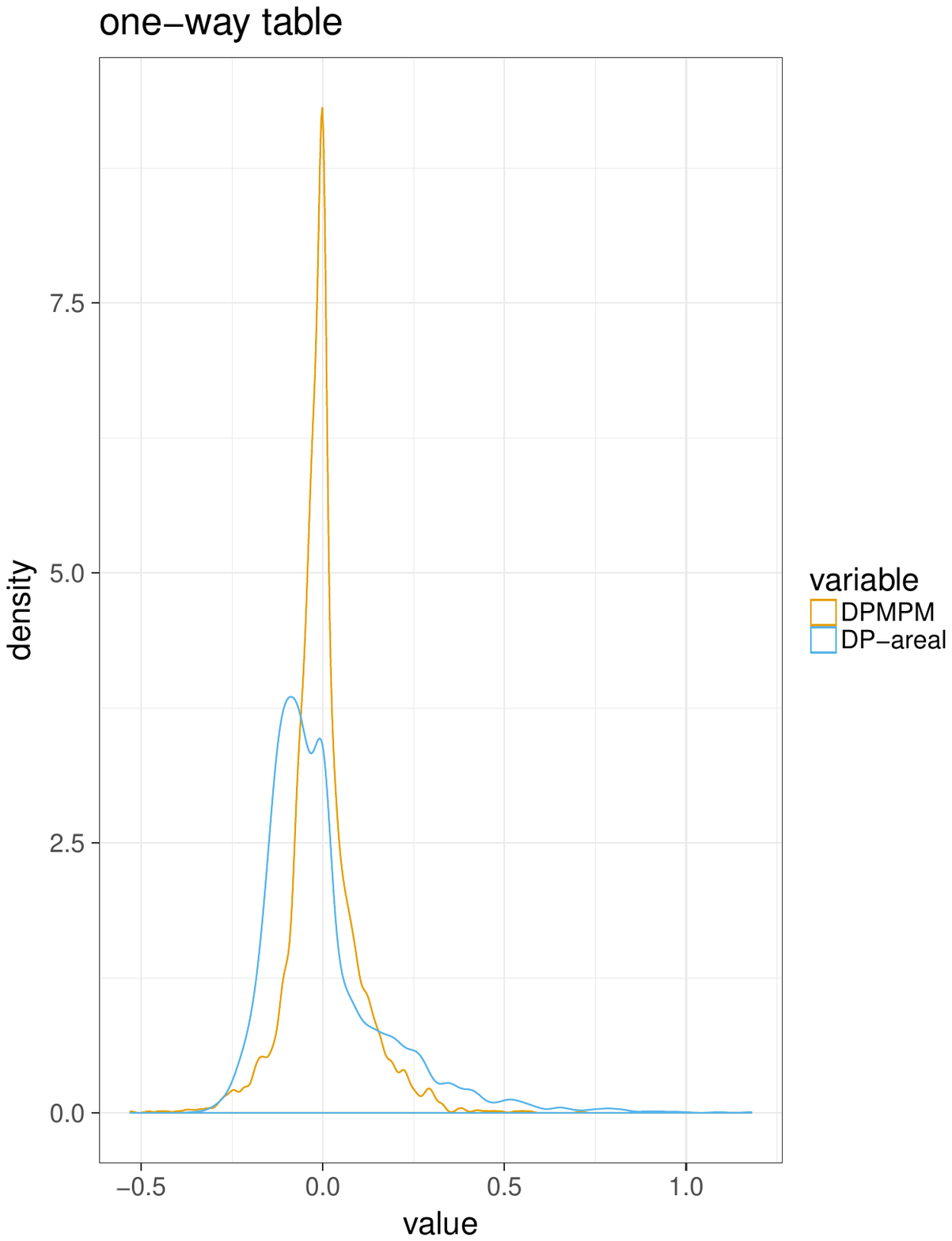}}
    ~
    \centering
    \subfloat[Two-way.\label{fig:twoway}]{
        \centering
        \includegraphics[width=0.33\textwidth]{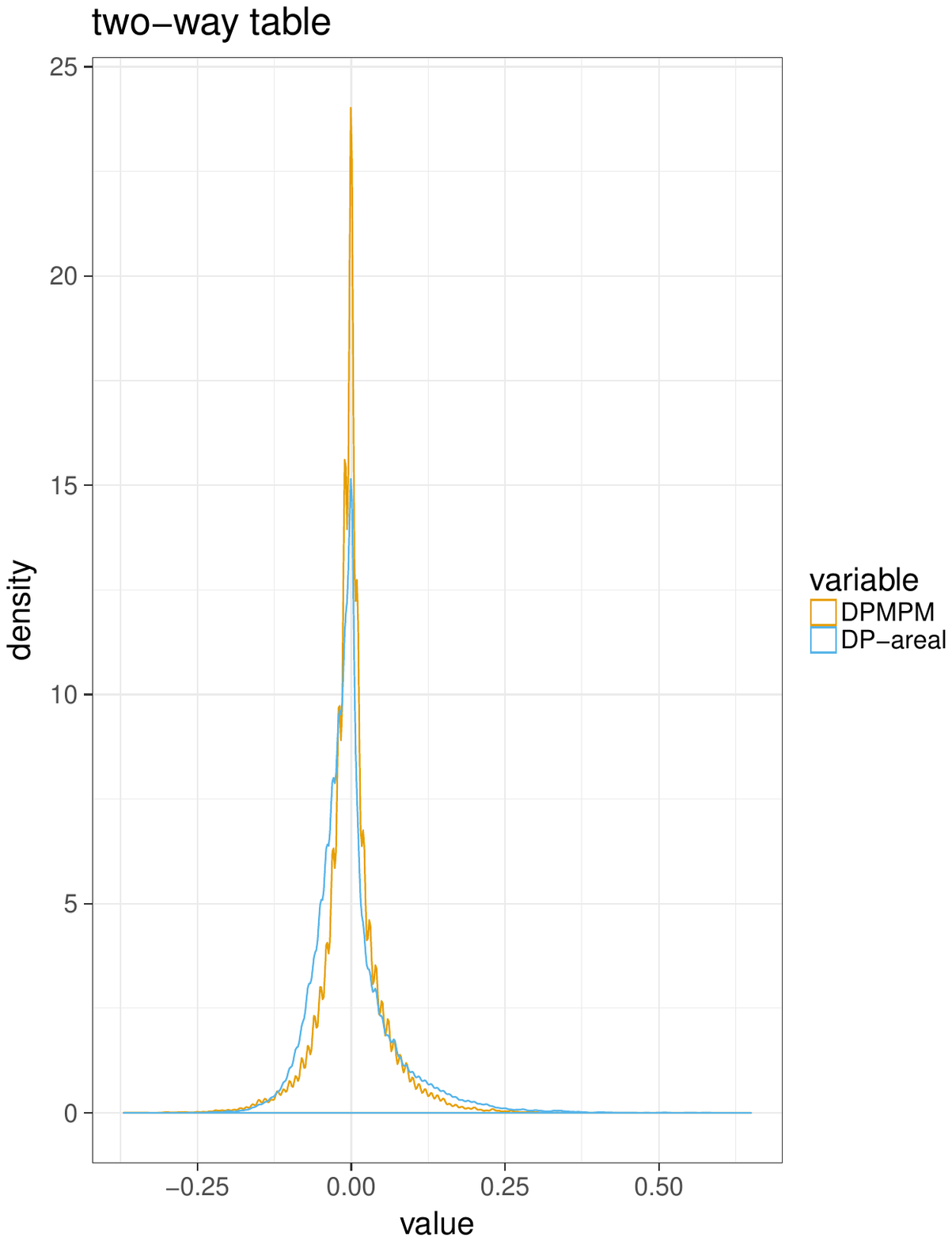}}
    ~
    \subfloat[Three-way.\label{fig:threeway}]{
        \centering
       \includegraphics[width=0.33\textwidth]{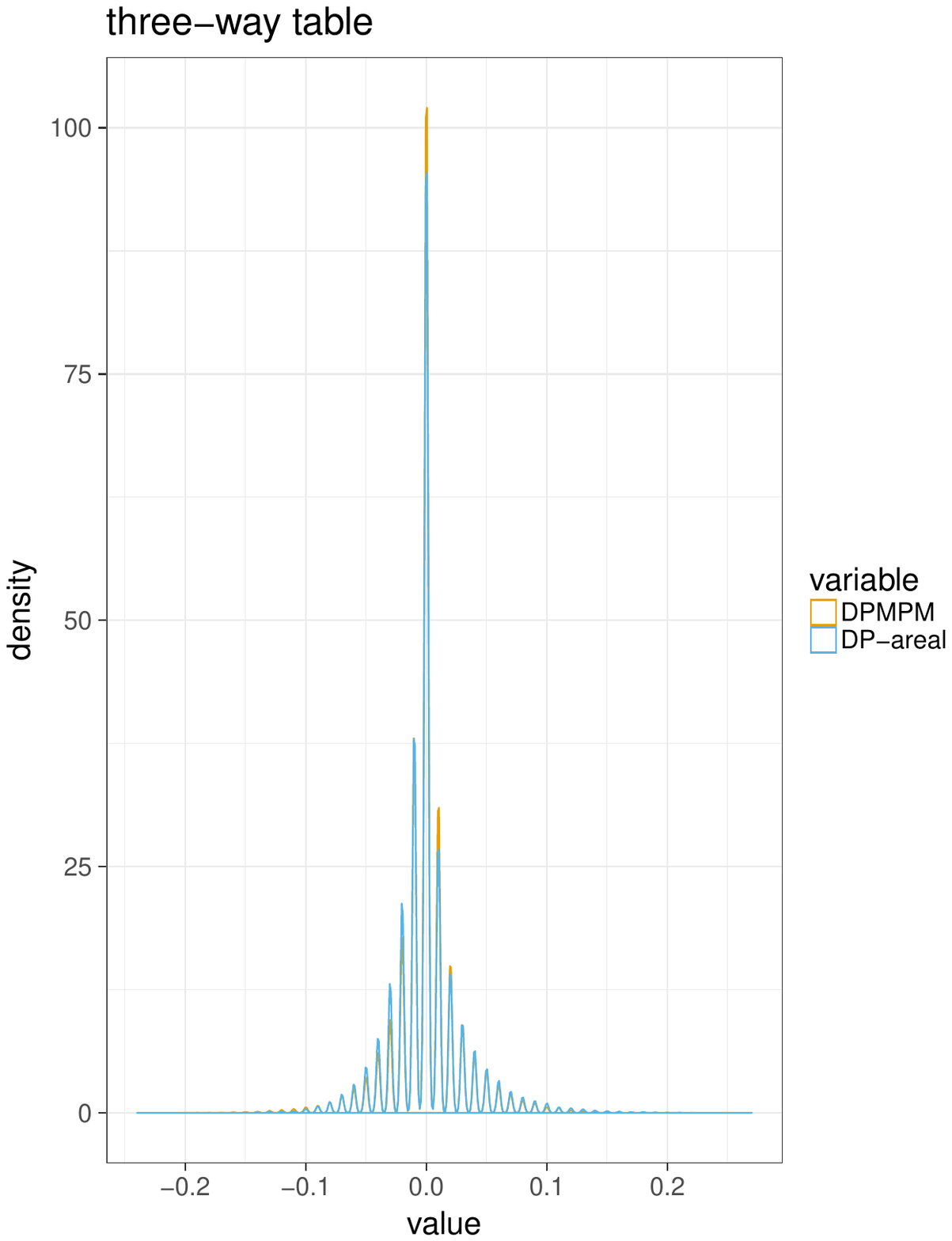}
}
    \caption{Distribution of actual deviations of one-way, two-way and three-way table counts produced by each synthesizer, DPMPM vs DP-areal, as compared to the original data over the $m = 20$ synthetic datasets.}
\end{figure}

Figure \ref{fig:oneway}, Figure \ref{fig:twoway}, and Figure \ref{fig:threeway} visualize the one-way, two-way, and three-way deviations in all $m = 20$ synthetic datasets through density plots. The actual deviations, not the absolute deviations, are plotted. If we focus on the x-axis of these plots, we can see that the range of actual count deviation of the synthetic data (from either synthesizer) from the original data is from -0.5 to 1 in one-way tables, -0.25 to 0.5 in two-way tables, and -0.2 to 0.2 in three-way tables. Since we are plotting the actual deviations over 133 counties and $m = 20$ synthetic datasets, such small ranges of deviations indicate high level of global utility preservation by both synthesizers: they have done a good job mixing CUs between counties while well-maintaining the distribution of counts among the counties. 

To compare the global utility preservation performance between our two synthesizers, we note that especially evident in Figure \ref{fig:oneway}, the one-way deviation in $\mathbf{Z}_{DPMPM}$ is more concentrated around 0 than that in $\mathbf{Z}_{DP-areal}$, indicating less overall deviation of the synthetic datasets generated by the DPMPM synthesizer from the original dataset. The findings are in accordance with Table \ref{tab:utility_DPMPMandDPareal}, showing higher level of preservation of distributional characteristics of county label (i.e. the utility of the synthetic data) by the DPMPM synthesizer than that by the DP-areal synthesizer. The DP-areal synthesizer, nevertheless, produces good utility. The density plots of actual deviations in two-way and three-way tables in Figure \ref{fig:twoway} and Figure \ref{fig:threeway} show better utility performance of the DPMPM synthesizer, overall, although the differences in performance is smaller compared to one-way tables. Overall, the global utility evaluation shows higher utility for synthetic data generated by the DPMPM synthesizer than that by the DP-areal synthesizer, though the utility of both synthesizers is good.

\subsection{Analysis-specific utility measure}
\label{utility:app}

We further assess the utility of the synthetic datasets by devising an inferential question that may be typical of the general type of analyses expected to be performed by the data analyst that go beyond table count queries.  Our application focuses on the prediction of income categories with an ordered logistic regression that regresses the CU's income category on the synthesized county label, age, and gender.  We use the \texttt{polr} function in \texttt{MASS} R package to fit the ordered logistic regression models.  We randomly select 5000 out of 6208 CU's to train the model and, subsequently, predict the expected income category of the remaining 1208 CU's.  This approach provides an out-of-sample assessment of the income prediction properties of the model using the county label variable.  We construct the predicted income for each data hold-out record by computing the expected value over the income categories using the predicted ordered category probabilities (that utilize regression coefficients estimated on the training data).  We compare how well predictions performed on the original, confidential data, on the one hand, accord with those performed on the $m = 20$ $\mathbf{Z}_{DPMPM}$ and $\mathbf{Z}_{DP-areal}$ synthetic datasets, on the other hand.

Figure \ref{fig:reg} presents the distribution (density) of the predicted expected income category of the 1208 CU's for each of the original data (grey curve), the DPMPM synthesizer (orange curve), and the DP-areal synthesizer (blue curve). The orange and blue curves are based on \emph{one} of $m = 20$ synthetic datasets generated by the two synthesizers respectively for readability and brevity.  Plots based on remaining synthetic datasets show similar results.  Overall we see a high level of preservation for the density by the two synthesizers, with the DPMPM synthesizer performing slightly better.

\begin{figure}[t]
\centering
\includegraphics[width=10cm, height=12cm]{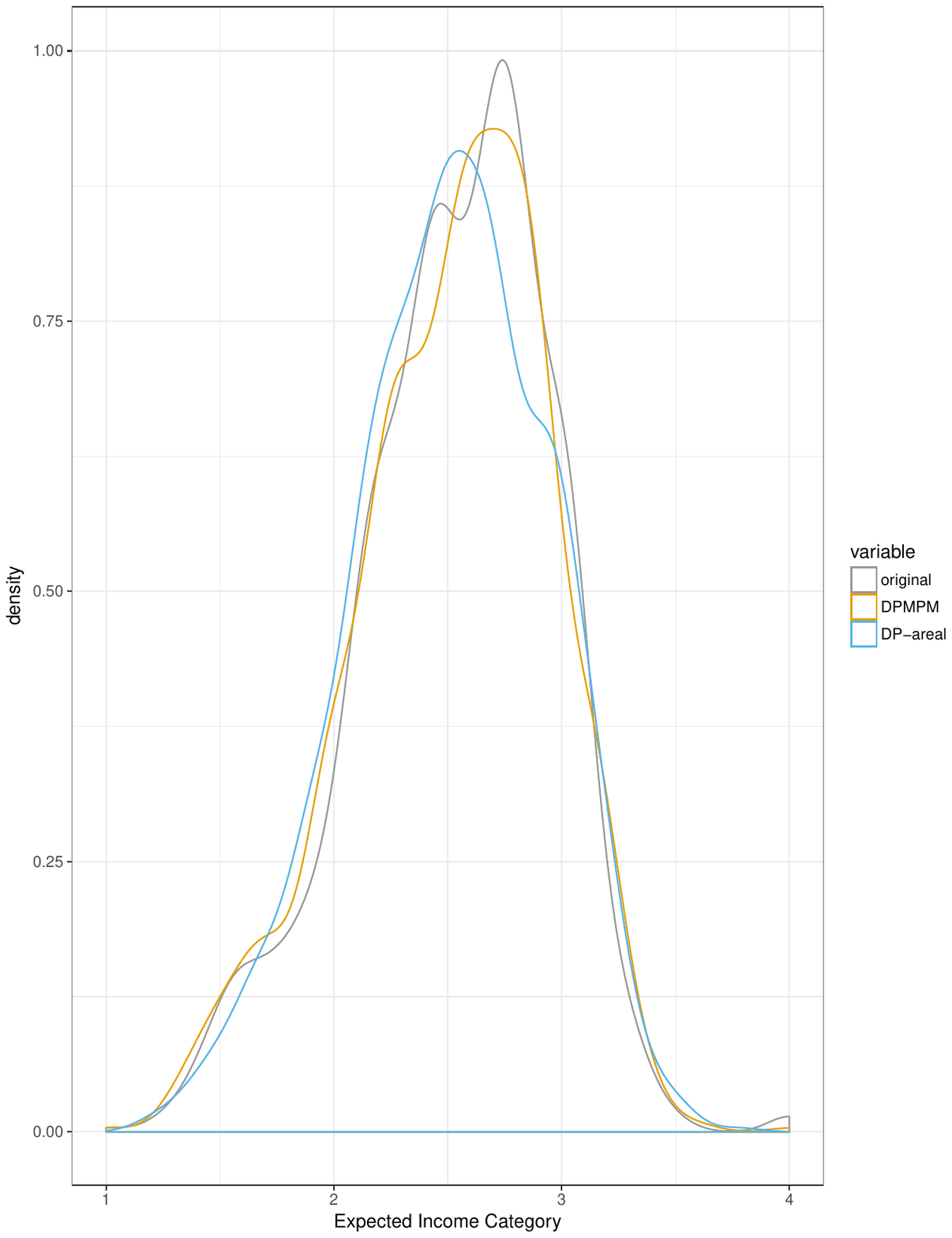}
\caption{Predicted expected income category in the original data, the $\mathbf{Z}_{DPMPM}$ synthetic datasets, and the $\mathbf{Z}_{DP-areal}$ synthetic datasets.}
\label{fig:reg}
\end{figure}

%

Next, we compare the uncertainty quantifications for a selection of county-indexed regression coefficients from our ordered logistic regression model estimated on the original, confidential data and on the $m = 20$ synthetic datasets for each of DPMPM and DP-areal.
In Table \ref{tab:reg}, we report 95\% confidence intervals of several representative regression coefficients from the model fits. Results in the Original Data column are obtained using the \texttt{confint.default} command of the model fit. Results in the DPMPM and DP-areal columns are obtained using simple combining rules over the $m = 20$ synthetic datasets for each synthesizer that account for \emph{both} the variation among the $m$ synthetic datasets, as well as the variation for the coefficients of interest within each dataset \citep{Drechsler2011book}.

As evident in Table \ref{tab:reg}, the confidence intervals and assessments of statistical significance are largely preserved in DPMPM and to a slightly lesser degree in DP-areal. The intervals are bit wider for DPMPM than the original data and DP-areal. The wider intervals for the synthetic data result from the greater uncertainty induced by the synthesizing process incorporated into combining rules \citep{Drechsler2011book}.

\begin{table}
\caption{Table of confidence intervals of representative regression coefficients, obtained from the same ordered logistic regression model fitted on the original data, fitted on the synthetic datasets $\mathbf{Z}_{DPMPM}$ ($m = 20$) and $\mathbf{Z}_{DP-areal}$ ($m = 20$), respectively. Results in the DPMPM and DP-areal columns are obtained using combining rules for partially synthetic datasets.\label{tab:reg}}
\centering
\begin{tabular}{c|lll}
\hline
 County & Original Data & DPMPM & DP-areal \\ \hline
 4 & [-0.546, -0.027] & [-0.682, -0.116] & [-0.593, 0.031] \\
 6 & [-0.301, 0.202] & [-0.426, 0.115] & [-0.344, 0.244] \\
 18 & [-0.516, -0.004] & [-0.661, -0.108]  & [-0.541, 0.059] \\
 32 & [-0.032, 0.476] & [-0.242, 0.295] & [-0.169, 0.413] \\
 33 & [-0.578, -0.065] & [-0.587, -0.045] & [-0.602, 0.004]\\ \hline
\end{tabular}
\end{table}


\subsection{Within pattern utility measure}
\label{utility:pattern}

Another approach for evaluating the synthetic data utility is to compare the induced distributions of the synthesized data records over the county labels within each pattern to the distribution for the original data. Recall that we define a pattern as a unique composition of non-geographic variables, as \{Gender, Income, Age\}, and there are 40 patterns in the CE data sample.  Our synthesizers are constructed as Bayesian hierarchical models where the prior distributions over parameters will induce ``smoothing" of the real data distribution in the resulting synthetic data by both smoothing over local features in the real data distribution and inducing distortion through shrinking.  The shrinking occurs under both synthesizers through the co-clustering of data records, which are then generated from the same distribution.  Since the partially synthetic CE datasets have only the county label synthesized, evaluating the preservation of distributional characteristics of county label within each pattern is of particular interest.  This utility measure is within pattern, or local.
\begin{figure}[t]
\centering
\includegraphics[width=12cm, height=12cm]{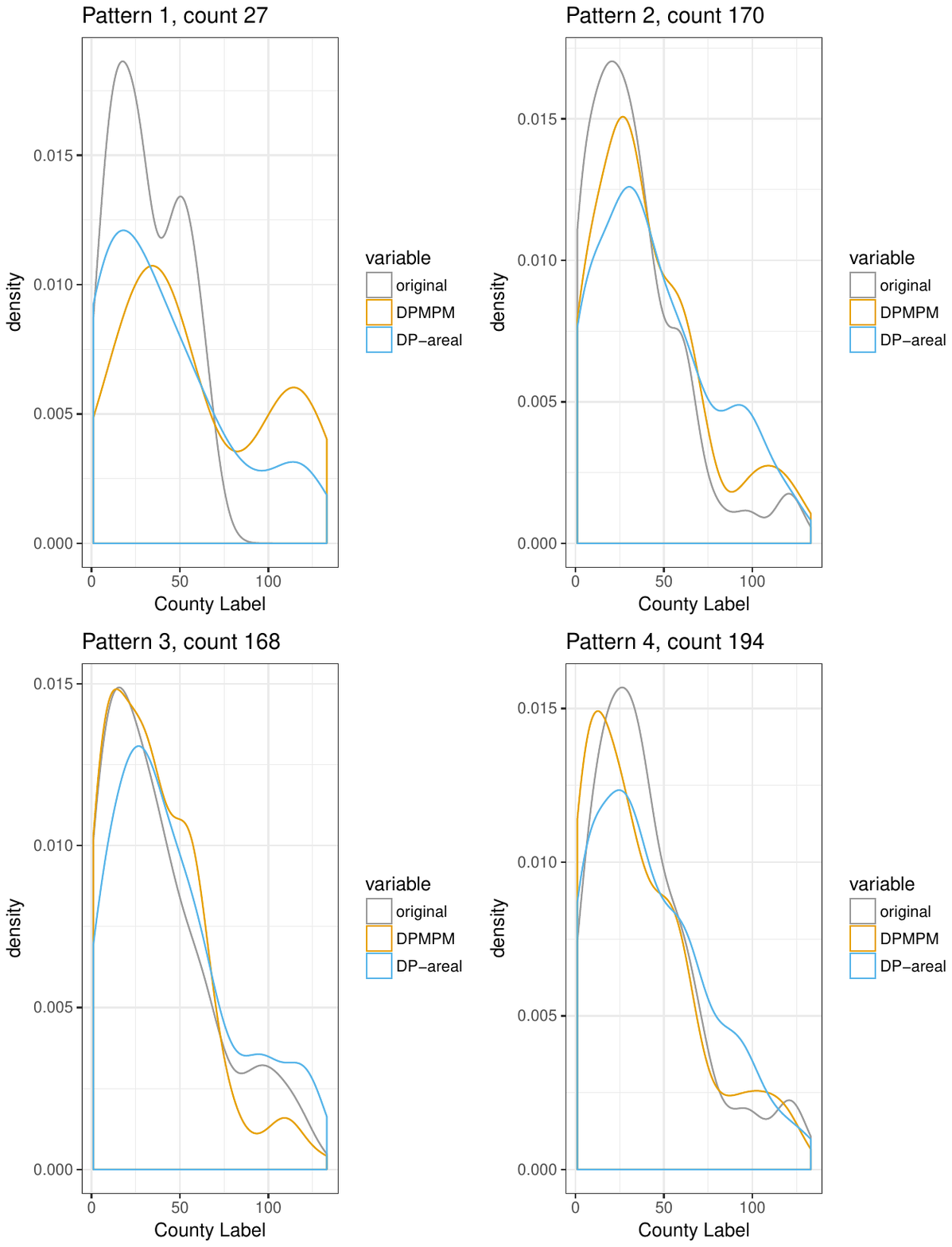}
\caption{Counties in Pattern 1 to Pattern 4.}
\label{fig:CountyLabelsPattern1to4}
\end{figure}

Each curve in each panel of Figure \ref{fig:CountyLabelsPattern1to4} presents a density approximation of a histogram (using the bandwidth of \citet{scott92}) of data records over the county labels for a chosen pattern.  The grey curve presents the \emph{original} data distribution of over the \emph{observed} county labels, the orange curve overlays the distribution over \emph{synthetic} county labels by the DPMPM synthesizer, and the blue curve overlays the distribution over synthetic county labels by the DP-areal synthesizer. The orange and blue curves are based on one of $m = 20$ synthetic dataset generated by the two synthesizers respectively for readability and brevity.  Plots based on remaining synthetic datasets show similar results. For brevity, density plots in Pattern 1 to Pattern 4 are included in the main text. The remaining 36 density plots are presented in the Supplementary Material.

Overall, in most of the patterns, the distribution of county label in synthetic datasets generated by the DPMPM synthesizer (orange curve) is closer to that of the original data (grey curve), than is the distribution generated by the DP-areal synthesizer (blue curve). The DPMPM better reproduces peaked behavior in the original data distribution, while the DP-areal model induces more smoothing.  The DPMPM also better reproduces local features in the county label distribution that are smoothed over by the DP-areal.   Both synthesizers, however, induce an equally high degree of smoothing or distortion in patterns with a small number of observations/records.

In summary, our DPMPM synthesizer better preserves the utility compared to the DP-areal synthesizer, for the various utility measures we have considered, although both synthesizers maintain high utility level. The DPMPM synthesizer directly models the county label as a categorical variable, while the DP-areal synthesizer indirectly models it through modeling the counts within each pattern. Moreover, the synthetic data generation process of the DPMPM synthesizer directly generates a synthesized value of county label from the fitted model, while that of the DP-areal synthesizer utilizes the estimated county-and-pattern specific probabilities to simulate a county label. It is the indirect modeling and synthesis approaches of the DP-areal synthesizer that negatively affects its utility performance.

\section{Disclosure risks measure}
\label{risks}

As we have seen in Section \ref{utility}, the DPMPM and the DP-areal synthesizers induce smoothing in the distribution of the county label as compared to the original data, both globally and within each pattern. The induced smoothing attempts to maintain the utility of the synthetic data, making it useful to data analysts for their analysis interests. At the same time, the induced smoothing provides privacy protection in the synthetic data. With county labels synthesized in the CE data, an intruder can no longer know the true category of the synthesized county label of any record. Moreover, she can no longer know the identification of any record with 100\% certainty, even though she could have access to the un-synthesized pattern of \{gender, age, income\} of that record. Nevertheless, an intruder could still make guesses about the true category of the synthesized county label, and make guesses about the identity of any record, using un-synthesized variables that might be available to her. The first type of risk, where the intruder seeks the value of the county label for a record, is commonly known as attribute disclosure.  The second type of risk, where the intruder seeks the identity of a record, is commonly referred to as identification disclosure.  We proceed to construct measures for attribute and identification disclosure risks.

Typically, both types of disclosure risks are measured for the simulated synthetic datasets, indicating level of protection (or the lack of) by the release of synthetic data. If multiple synthesizers have been proposed, as in our current application for synthetic CE data, evaluations of disclosure risks measures and their comparisons could inform the CE program about which synthesizer provides higher privacy protection. In the final analysis, the data disseminating agencies, such as the CE program, are able to make a decision among the synthesizers through evaluating their relative usefulness (i.e. utility measures) and the level of privacy protection that they encode (i.e. the disclosure risk measures). If disclosure risk measures can be developed for the original data, agencies will have much more information when deciding among synthesizers based on not only comparison of their disclosure risk measures to each other, but also on comparison to those in the original data.
We next describe our general approach to measure disclosure risks in the original data.

\subsection{Inherent disclosure risks in the CE data}
\label{risks:inherent}

%

A synthesizer replaces the county label for every record in the original data with a synthesized value.  In the limit as a synthesizer becomes more accurate, the best synthesizer may be imagined as one that generates the synthesized values of the county label for all records by using the \emph{original} data distribution.  We label this type of synthesizer that uses the original data distribution as ``perfect".  Generating synthetic values under the original data distribution is both independent of any synthesizing model and may produce synthesized values that differ from the original data.  A perfect synthesizer provides the highest possible utility, because there is no deviation from the original data distribution; however, it increases disclosure risks at the same time. Although the original data is not released to the public, as data disseminators, we can use it to construct maximum disclosure risks measures and create an upper bound of the acceptable disclosure risks.

To mimic the behavior of an intruder with the highest amount of information (i.e. the exact distribution of county label), we can sample a new draw of the county label for every record based on its original distribution within each pattern.  We approximate the distribution of county labels in the original data with the empirical distribution, which we sample under a weighted re-sampling scheme.  Given this set of newly sampled county labels, we can then calculate the identification disclosure risks and attribute disclosure risks. We repeat this process for a large number of times, and obtain sampling distributions of two types of disclosure risks, fully capturing the variability in the sampling processes. Disclosure risks computed based on this repeated sampling procedure provide an upper bound of the acceptable disclosure risks, because this procedure uses maximum amount of information that may be published - the original data distribution. Therefore, we label this scenario as the maximum disclosure risks scenario.

By contrast, we may establish a lower bound on the risk that may be achieved by a synthesizer.   This requires us to go to the other extreme, where an intruder has the \emph{least} amount of information about the distribution of the county label in the original data.  We employ a uniform distribution over among all possible county labels within each pattern as the minimally-informative scenario; i.e., we can sample a new draw of the county label of every record from a uniform distribution over the 133 observed county labels within each pattern in the CE data sample. Given this set of newly sampled county labels, identification and attribute disclosure risks can be calculated, and this process is repeated for a large number of times to obtain the sampling distributions of two types of disclosure risks.  Because this repeated sampling procedure uses the minimum amount of information, disclosure risks computed based on this procedure provide a lower bound of the acceptable disclosure risks, and we label this scenario as the minimum disclosure risks scenario.  Similar to the maximum disclosure risks scenario, the minimum scenario is a type of risk that is inherent in the original dataset and independent of the synthesizer.

\subsection{Identification disclosure risks}
\label{risks:identification}

\subsubsection{Three summaries of identification disclosure probabilities}
Identification disclosure risks measure how likely it is for an intruder to correctly identify a record by matching with available information from external files. In our current application, the released synthetic datasets contain three un-synthesized variables (gender, age, and income), and one synthesized variable (county label). Suppose an intruder has access to an external file that includes gender, age, and county label of every record, as well as their identities.  The attribute values, but not the identity of the records, also appear in the released synthetic datasets. With access to such external information, the intruder may attempt to identify a record by performing matches within each pattern.  The matching is performed within pattern because the intruder knows the values of the pattern attribute value are not synthesized.

Without loss of generality, assume that the intruder has external information about every record's gender, age, and county label. Let $c_i$ be the number of records with the highest match probability for record, $i$ (i.e. the number of records having the \emph{exact} same values of gender, age, and county label as record $i$ in the original data); let $T_i = 1$ if the true match is among the $c_i$ units and $T_i = 0$, otherwise. We recall that $T_{i}$ will be set to $1$ if the synthesized value for the county label for record, $i$, is the same as that for the original data.  Let $K_i = 1$ when $c_i T_i = 1$ and $K_i = 0$ otherwise (i.e., $K_i = 1$ indicates a true unique match exists), and let $n$ denote the total number of target records (i.e. every record in the CE data). Finally, let $F_i = 1$ when $c_i (1 - T_i) = 1$ and $F_i = 0$ otherwise (i.e., $F_i = 1$ indicates a false unique match), and let $s$ equal the number of records with $c_i = 1$ (i.e. the number of records that are uniquely matched among $n$ target records). There are three widely used file-level summaries of identification disclosure probabilities using the notations and definitions given above \citep{ReiterMitra2009, DrechslerReiter2010, HuHoshino2018PSD, Hu2018TDP, DrechslerHu2018}.
\begin{enumerate}
\item[(a)] The expected match risk:
\begin{eqnarray}
\sum_{i=1}^{n} \frac{T_i}{c_i}.
\end{eqnarray}
When $T_i = 1$ and $c_i > 1$, the contribution of unit $i$ to the expected match risk reflects the intruder randomly guessing at the correct match from the $c_i$ candidates, where the intruder probability of a correct guess is $1/c_{i}$. In general, the higher the expected match risk, the higher the identification disclosure risks.

\item[(b)] The true match rate:
\begin{eqnarray}
\sum_{i=1}^{n} \frac{K_i}{n},
\end{eqnarray}
which is the percentage of true unique matches among the target records. In general, the higher the true match rate, the higher the identification disclosure risks.

\item[(c)] The false match rate:
\begin{eqnarray}
\sum_{i=1}^{n} \frac{F_i}{s},
\end{eqnarray}
which is the percentage of false matches among unique matches. In general, the lower the false match rate, the higher the identification disclosure risks.
\end{enumerate}


\subsubsection{Results of identification disclosure risks}

When performing matching with external files, there are different assumptions about the intruder's knowledge of the un-synthesized variables. We consider two cases of assumption of intruder's knowledge, encoded in column Known in Table \ref{tab:IdRisks_DPMPMandDPareal}:  i) only gender and county label; ii) gender, age, and county label.  We collapse across combinations of un-synthesized variables in case ii) to achieve case i).  We would expect identification risks to be generally lower as we collapse across combinations since $c_{i}$ would be expected to increase. Such is not always the case, however, as we observe for the DP-areal results, presented below.

For each of the two cases, we report summaries of expected risk, true match rate, and false match rate of identification disclosure risks. The column DPMPM reports average summaries of $m = 20$ synthetic datasets, $\mathbf{Z}_{DPMPM}$ generated by the DPMPM synthesizer. The column DP-areal reports average summaries of $m = 20$ synthetic datasets, $\mathbf{Z}_{DP-areal}$ generated by the DP-areal synthesizer. The column Max reports average summaries of $S = 100$ repeated sampling under the maximum identification disclosure risks scenario. We exclude the column Min for brevity, and it has 0 for expected risk and true match rate for all cases, and NaN for false match rate for all cases (due to $s = 0$, i.e. no unique matches, in the denominator).

A subtle point about the Max for the false match rate is that because of the definition of the false match rate (the percentage of false matches among unique matches; and higher false match rate means higher privacy protection), the computed identification risk measures in the Max column is actually the lower bound of the acceptable range of false match rate; however, the Max for the expected risk and the true match rate serve as upper bounds, as discussed before.

\begin{table}
\caption{Expected risk, true match rate, and false match rate of identification disclosure risks of the synthetic datasets. Results are averages of $m = 20$ partially synthetic datasets for the columns DPMPM and DP-areal, and averages of $S = 100$ repeated sampling iterations for the Min column. (*computed based on non $s = 0$ cases) \label{tab:IdRisks_DPMPMandDPareal}}
\centering
\begin{tabular}{l|r|r|r|r}
\hline
 Known & Summary &  DPMPM & DP-areal&Max\\ \hline
 gender, and & expected match risk &  2.497 & 0.952  & 4.708 \\
  county label& true match rate &0.000  & 0  &0.000 \\
 & false match rate &0.997 & 1*  &  0.989\\ \hline
 gender,age, & expected match risk &10.474   &0.851 &  20.073 \\
  and county label& true match rate & 0.000 &0.000  &  0.001 \\
  & false match rate &0.991 & 1.000 &  0.978 \\ \hline
\end{tabular}
\end{table}

As evident in Table \ref{tab:IdRisks_DPMPMandDPareal}, for every case of assumption of intruder's knowledge, summaries of identification disclosure risks indicate significantly lower expected risk in synthetic data generated by the DP-areal synthesizer. As the intruder's knowledge increases, the expected risk in the DPMPM synthetic data increases and approaches to the Max, while the expected risk in the DP-areal synthetic data slightly decreases. On average, expected risk in the DPMPM synthetic data and the DP-areal synthetic data is bounded below by Min and bounded above by Max. The results of the true match rate and the false match rate show much smaller difference between the two synthesizers and between each synthesizer with the Max. Overall the DPMPM synthesizer and the DP-areal synthesizer has 0 or close to 0 true match rate, and 1 or close to 1 false match rate, suggesting high level of privacy protection.

To take a closer look at the expected risk, consider case ii) where exact matching is done assuming intruder's knowledge of gender, age, and county label. Recall that there are $n = 6208$ observations in the CE sample. For the DPMPM synthetic datasets, an average 10.474 expected risk indicates a record-level average 0.00169 expected risk when dividing by $n$. The corresponding record-level average expected risk from the DP-areal synthetic datasets is 0.00014, and that from the $S = 100$ repeated sampling under the maximum risk scenario is 0.00323. Overall, the expected risk is very low with both synthesizers, and it is low even under the maximum risk scenario, which suggests low inherent identification disclosure risks in the original CE data.

To visualize the variability among the summaries of $m = 20$ synthetic datasets for each of $\mathbf{Z}_{DPMPM}$ and $\mathbf{Z}_{DP-areal}$, Figure \ref{fig:IdMaxHist2} presents a histogram of $S = 100$ repeated samples of the expected risk under the maximum disclosure risks scenario is plotted. In addition, the minimum, mean, and maximum expected risk among $m = 20$ DPMPM synthetic datasets (dashed and orange) and those among $m = 20$ DP-areal synthetic datasets (dotted and blue) are also co-plotted.  Hidden in the average summaries of expected risk in Table \ref{tab:IdRisks_DPMPMandDPareal} is that although the average expected risk among $m = 20$ synthetic datasets $\mathbf{Z}_{DPMPM}$ is as small as half of the average upper bound formed by the average expected risk among $S = 100$ repeated samples under the maximum disclosure risks scenario, the expected risk computed for the DPMPM synthetic dataset (dashed and orange) shows close distance to the expected risk computed for repeatedly sampled ``synthetic" datasets under the maximum disclosure risks scenario (filled histogram). The maximum expected risk among $m = 20$ DPMPM synthetic datasets appears almost as large as the smallest expected risk among the repeated samples of ``synthetic" datasets under the maximum disclosure risks scenario, which may be cause for concern about DPMPM synthetic datasets. By contrast, the variability in expected risk among $m = 20$ DP-areal synthetic datasets (dotted and blue) is much smaller, and the expected risk for the DP-areal synthetic dataset is overall much smaller than those computed under the maximum disclosure risks scenario, as shown in Table \ref{tab:IdRisks_DPMPMandDPareal}, suggesting acceptable identification disclosure risks in DP-areal synthetic datasets.
A similar plot for case i) gender and county label of intruder's knowledge assumptions suggest overall acceptable identification disclosure risks for both the DPMPM and the DP-areal synthesizers, which agree with the average summaries in Table \ref{tab:IdRisks_DPMPMandDPareal}. The plot is included in the Supplementary Material for brevity.


\begin{figure}[t]
\centering
    \subfloat[Expected identification disclosure risks. Histogram of expected risks under the maximum risk scenario. Known variables: gender, age, and county label.\label{fig:IdMaxHist2}]{
        \centering
        \includegraphics[width=0.4\textwidth]{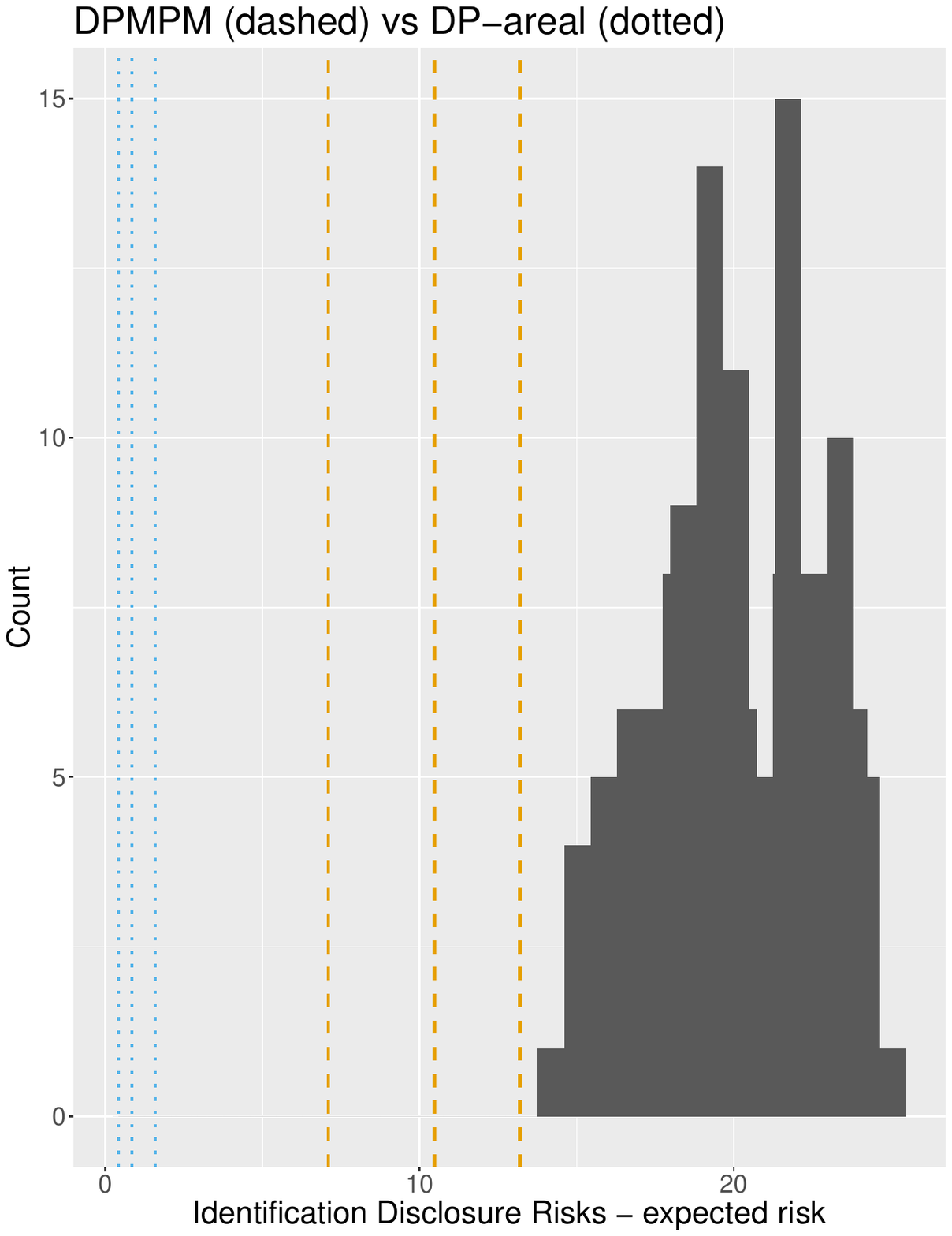}}
    ~
    \centering
    \subfloat[Exact attribute disclosure risks. Histograms of exact attribute disclosure risks under the minimum and the maximum risk scenarios. Known variables: gender, age, and income. \label{fig:AttMaxMin}]{
        \centering
        \includegraphics[width=0.4\textwidth]{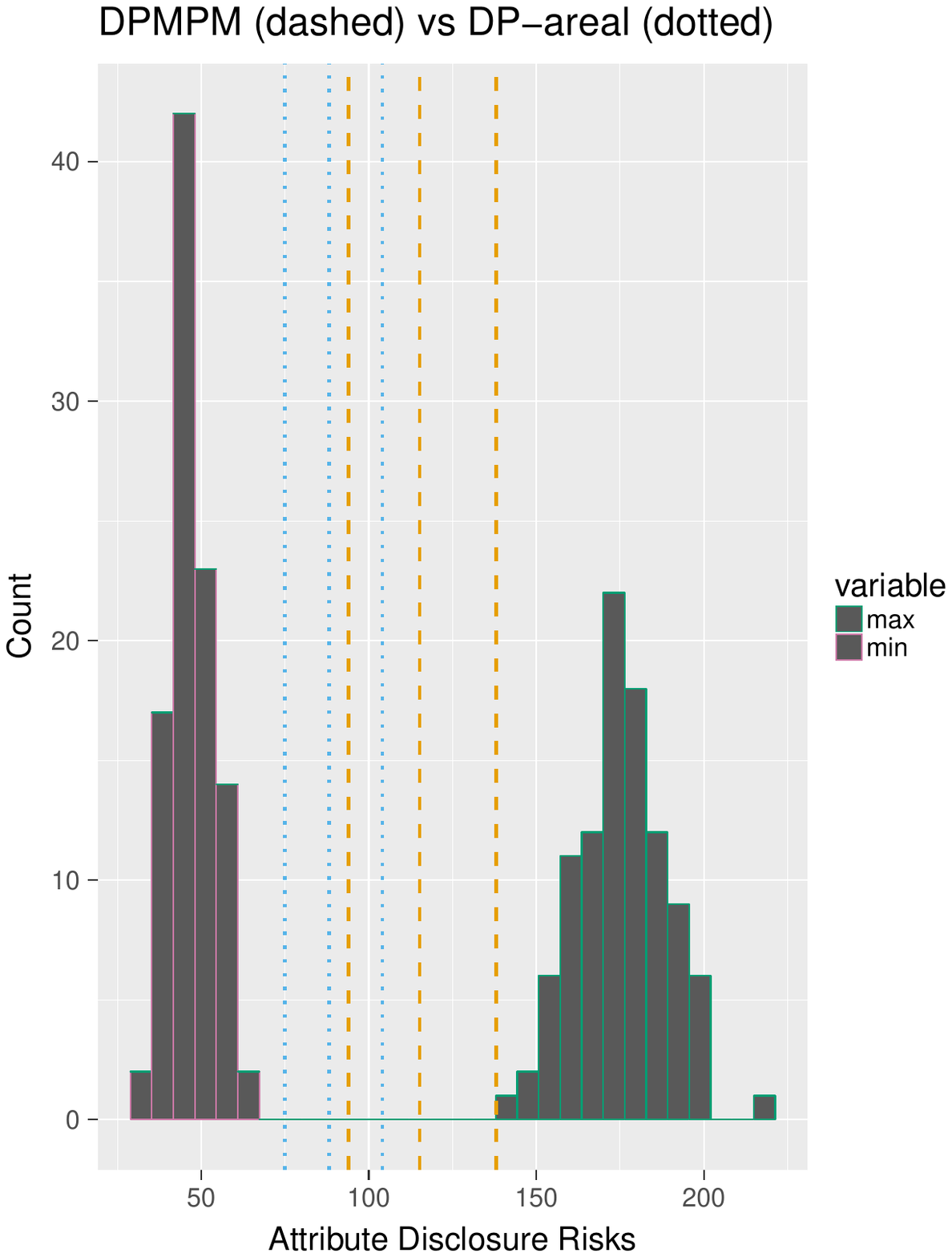}
}
    \caption{Histograms of disclosure risks under the maximum risk scenario and/or the minimum risk scenario. Vertical lines include the min, mean, and max among the $m = 20$ synthetic datasets.}
\end{figure}

\subsection{Attribute disclosure risks}
\label{risks:attribute}

\subsubsection{The summary of exact attribute disclosure risks}
Attribute disclosure risks measure how likely it is for an intruder to correctly infer the true value of a synthesized variable or attribute in the original data from the synthetic datasets. Such attacks usually make use of all un-synthesized attributes, therefore we only consider the case where the intruder uses the available pattern of each record (i.e. gender, age, and income), to infer the true county label in the original dataset. That is, we assume gender, age, and income are available when mimicking the intruder's behavior to conduct attribute attacks. Let $A_i = 1$ if the synthesized county label category is the same as the original county label for record $i$, and $A_i = 0$, otherwise (i.e., $A_i = 1$ indicates an exact attribute disclosure). The number of exact attribute disclosures is
\begin{equation}
\sum_{i = 1}^{n} A_i.
\label{eq:ExactAttributeDisclosure}
\end{equation}
We note that some attribute disclosure risks for variables containing geographic information proposed in previous works focus on distance between the synthesized location and the true location. For every record, \citet{WangReiter2012} reported a Euclidean distance $R_1$ between the intruder's guess of the longitude and latitude and the actual longitude and latitude, and then reported the count $R_2$ recording the number of actual cases in circle centered at the actual longitude and latitude with radius $R_1$. \citet{Paiva2014SM} reported a Euclidean distance measure between the true location $\mathbf{y}_i$ and the guess $\mathbf{y}^{\ast}$ with the maximum posterior probability of record $i$.
Because the county label in the CE data sample is treated as categorical, and moreover because of little geographic information this variable carries, as discussed in Section \ref{intro:CEdata}, measuring a Euclidean distance between the true county label and the synthesized county label for record $i$ is not feasible. Therefore, we only consider the number of exact attribute disclosures in Equation (\ref{eq:ExactAttributeDisclosure}). Our measure for attribute risk is composed as the sum of exact attribute disclosures for all records, and an exact attribute disclosure for record $i$ is declared when the synthesized county label category is the same as the county label category in the original data.  We construct our definition for attribute risks to be consistent with that for identification risks, where both produce file-level measures of risk designed to help reporting agencies assess the overall risk associated with the potential release of the synthetic data.  

\subsubsection{Results of attribute disclosure risks}

The first row of Table \ref{tab:AttRisksDPMPMandDPareal} presents the average numbers of exact attribute disclosures among $m = 20$ DPMPM synthetic datasets $\mathbf{Z}_{DPMPM}$, and the same for $\mathbf{Z}_{DP-areal}$, in the DPMPM and DP-areal columns, respectively. It also reports the average numbers of exact attribute disclosures among $S = 100$ repeated sampling iterations under the minimum disclosure risks scenario, and among $S = 100$ repeated sampling iterations under the maximum disclosure risks scenario, in the Min and Max columns respectively. The second row presents corresponding percentages of exact attribute disclosures, by dividing the values in the first row with $n = 6208$.  These results show that, on average, the numbers of exact attribute disclosures in both the DPMPM synthetic data and in the DP-areal synthetic data are generally far away from the maximum scenario, with the latter lower than the former, consistent with results for identification disclosures risk.  The inherent risks in the original data are not high when given the maximum amount of information, while it is not zero when given the minimum amount of information.

\begin{table}
\caption{The average numbers and percentages of exact attribute disclosures. Results are averages of $m = 20$ partially synthetic datasets for the columns DPMPM and DP-areal, and averages of $S = 100$ repeated sampling iterations for the Min and Max columns.\label{tab:AttRisksDPMPMandDPareal}}
\centering
\begin{tabular}{r|r|r|r|r}
\hline
 Summary &  Min & DPMPM & DP-areal&  Max\\
 \hline
 Number of exact attribute disclosures & 47.25 & 115.25 & 88.15 & 175.33\\
 Percentage of exact attribute disclosures & 0.76\% & 1.86\% & 1.42\% & 2.82\% \\
 \hline
\end{tabular}
\end{table}


Figure \ref{fig:AttMaxMin} plots a histogram of exact attribute disclosures based on $S = 100$ iterations of repeated sampling under the minimum disclosure risks scenario (purple), and another histogram under the maximum disclosure risks scenario (green). Additionally, vertical lines of the results from the DPMPM synthetic datasets (dashed and orange) and from the DP-areal synthetic datasets (dotted and blue) are plotted for comparison. Each set of three vertical lines correspond to the minimum, the mean, and the maximum of the number of exact attribute disclosures among the $m = 20$ synthetic datasets.  Figure \ref{fig:AttMaxMin} agrees with the results in Table \ref{tab:AttRisksDPMPMandDPareal}, showing that when considering the variability in sampling, the DP-areal synthetic datasets have generally lower attribute disclosure risks compared to the DPMPM synthetic datasets, and both are well below the maximum disclosure risk scenarios.

In summary, while our DPMPM synthesizer better preserves the utility, it also carries higher disclosure risks, both identification disclosure risks and attribute disclosure risks, compared to the DP-areal synthesizer. As discussed at the end of Section \ref{risks:identification}, the indirect modeling and synthesis approaches of the DP-areal synthesizer decrease its utility with higher level of smoothing, compared to the DPMPM synthesizer. It is this higher level of smoothing that provides higher level of privacy protection, resulting in lower disclosure risks carried by the DP-areal synthesizer. 

\section{Conclusion}
\label{conclusion}
We devised an end-to-end process for data synthesis, including formulating data synthesizers and measuring and comparing their utilities and disclosure risks in a fashion, that promotes ease-of-interpretation to facilitate decision making by statistical agencies who may consider the release of respondent-level synthetic data.  Our data synthesizers are constructed for the challenging case of generating geographic location labels.  Our formulations replace spatial priors with more general nonparametric prior formulations due to geographic sparsity, with the result that they may broadly apply to the synthesis of any polytomous variable characterized by multiple levels.  We leveraged the patterns formed from the intersection of known categorical variables to define a new local utility measure based on the distribution of the synthesized county label that makes intuitive the comparisons of usefulness among the synthesizers.   We designed new minimum and maximum risk measures that characterize the data and are independent of the choice of synthesizer.  These minimum and maximum risk bounds set context for evaluating the relative improvements in privacy protection provided by the synthesizers in a way that aids the agency decision-maker to evaluate whether synthetic data is sufficiently privacy protected for release.

Our applications to the CE data sample have shown that the DPMPM synthesizer, compared to the DP-areal synthesizer, produces synthetic datasets with higher utility, although such higher utility performance comes at the price of higher disclosure risks. Such utility-risk trade-offs are important evaluation criteria for statistical agency to make decisions about synthetic data dissemination. We advocate for the evaluation of inherent risks of the original, confidential data, as we have done in our applications and evaluations. The minimum and maximum risk bounds of the original data can greatly facilitate agencies' decision making process. 

Plans of releasing the synthetic CE county label are under way. With our synthetic CE datasets providing high utility and low disclosure risks, the CE survey program is making plans to produce synthetic CE datasets as a part of their experimental research products. Examples of their current experimental research products include state weight files.

An opportunity for future work is to construct a simple mechanism that regulates the amount of smoothing produced by a synthesizer to allow exploration of the utility-risk trade-offs. Future directions of applications include whether a negative binomial regression is better suited than a Poisson as in the DP-areal model, assessing utility-risk trade-offs when more variables are synthesized.

\section*{Acknowledgement}

This research is supported by ASA/NSF/BLS Senior Research Fellow Program. The authors thank Adam Safir, Steve Henderson, Geoffrey Paulin, Arcenis Rojas, Clayton Knappenberger, Taylor Wilson, and Daniel Yang at the U.S. Bureau of Labor Statistics, for thoughtful discussions and valuable suggestions.

\bibliographystyle{natbib}
\bibliography{CEbib}

\end{document}


\title{\large \bf  Supplementary Material to Bayesian Data Synthesis and Disclosure Risk Quantification: An Application to the Consumer Expenditure Surveys}

\author{Jingchen Hu\footnote{Vassar College, Box 27, 124 Raymond Ave, Poughkeepsie, NY 12604, jihu@vassar.edu.} $\,$ and Terrance D. Savitsky\footnote{U.S. Bureau of Labor Statistics, Office of Survey Methods Research, Suite 5930, 2 Massachusetts Ave NE Washington, DC 20212, Savitsky.Terrance@bls.gov.}}

\maketitle

\begin{abstract}
This supplement contains:  1. The list of 40 patterns in the CE sample; 2. The full set of within pattern distribution plots of the county label synthesized from the DPMPM, DP-areal synthesizers and the original data distribution; 3. A histogram of identification disclosure risks for the DPMPM, DP-areal and Maximum (from the original data) for the case where only gender and county label are known by the intruder; 4. The Stan script to implement the DP-areal synthesizer.

{\bf keywords}: Data privacy protection, Disclosure risks, Identification risks, Attribute risks, Synthetic data, Bayesian hierarchical models

\end{abstract}

\newpage
\section{List of 40 Patterns}
Table \ref{tab:patterns} lists the 40 patterns in the CE sample.
\vspace{1mm}
\begin{table}
\caption{List of 40 patterns: the index and the number of observations in each pattern.\label{tab:patterns}} 
\centering
\begin{tabular}{rr|rr}
\hline
 Index &  Observations & Index &  Observations \\ \hline
 1 & 27 & 21 & 33 \\
 2 & 170 & 22 & 229 \\
 3 & 168 & 23 & 222 \\
 4 & 194 & 24 & 333 \\
 5 & 48 & 25 & 128 \\
 6 & 3 & 26 & 9 \\
 7 & 193 & 27 & 250 \\
 8 & 183 & 28 & 254 \\
 9 & 242 & 29 & 308 \\
 10 & 61 & 30 & 53 \\
 11 & 3 & 31 & 8 \\
 12 & 291 & 32 & 244 \\
 13 & 275 & 33 & 312 \\
 14 & 199 & 34 & 184 \\
 15 & 19 & 35 & 18 \\
 16 & 4 & 36 & 3 \\
 17 & 239 & 37 & 198 \\
 18 & 454 & 38 & 344 \\
 19 & 169 & 39 & 122 \\
 20 & 4 & 40 & 10 \\ \hline
\end{tabular}
\end{table}

\newpage

\section{Within Pattern Density Plots of County Labels among the Synthesizers}

\begin{figure}[H]
\centering
\includegraphics[width=12cm, height=12cm]{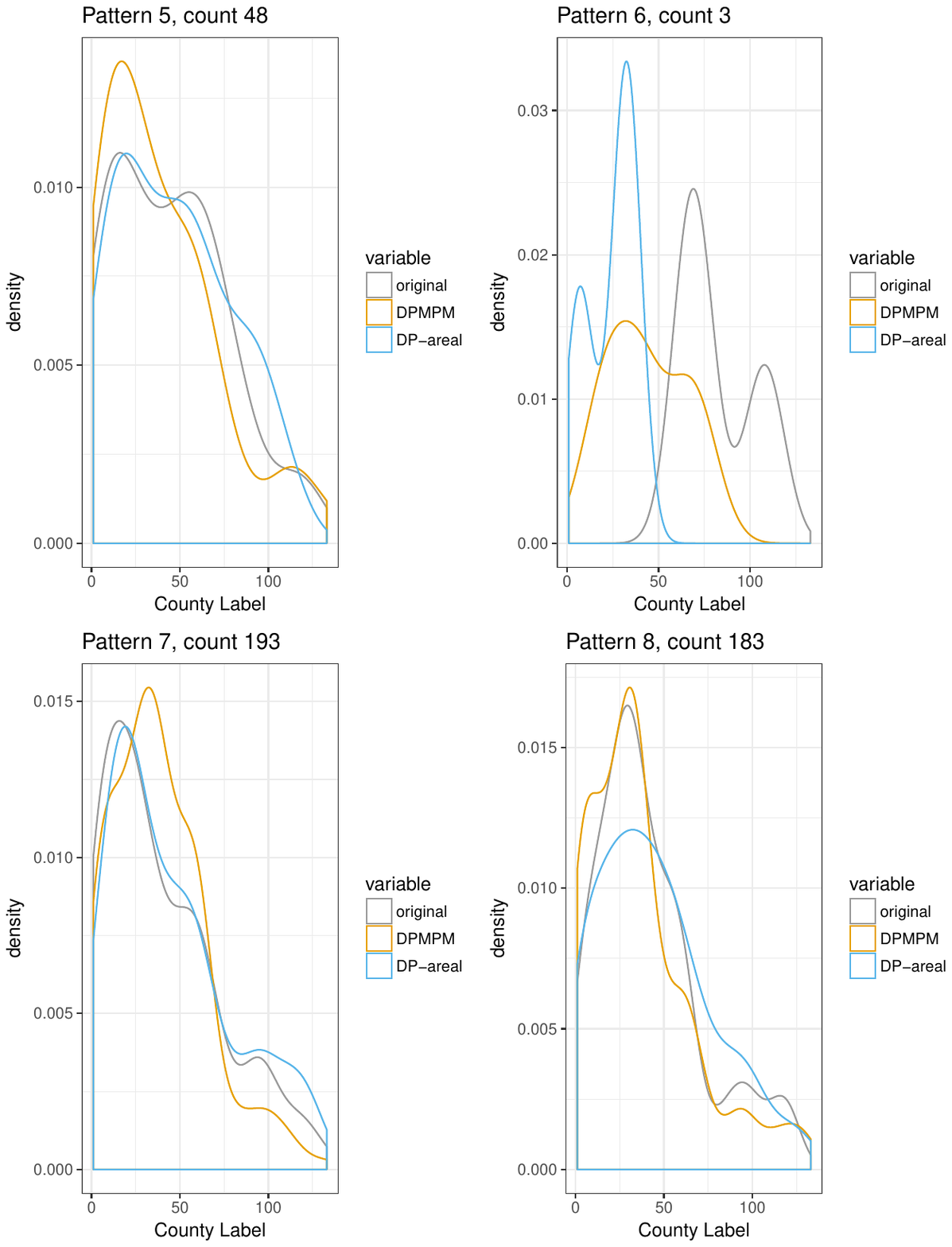}
\caption{Counties in Pattern 5 to Pattern 8.}
\label{fig:CountyLabelsPattern5to8}
\end{figure}

\begin{figure}[H]
\centering
\includegraphics[width=12cm, height=12cm]{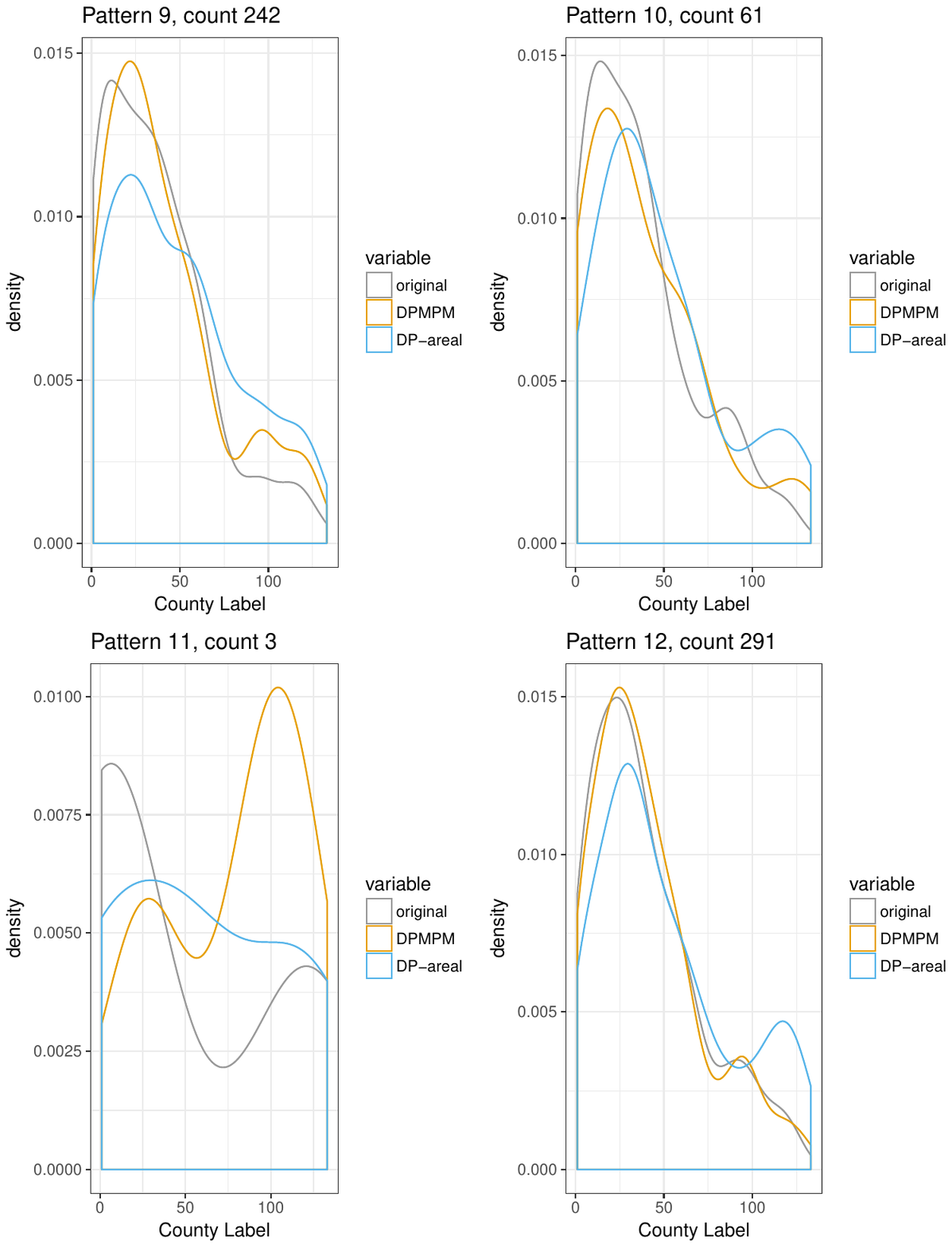}
\caption{Counties in Pattern 9 to Pattern 12.}
\label{fig:CountyLabelsPattern9to12}
\end{figure}

\begin{figure}[H]
\centering
\includegraphics[width=12cm, height=12cm]{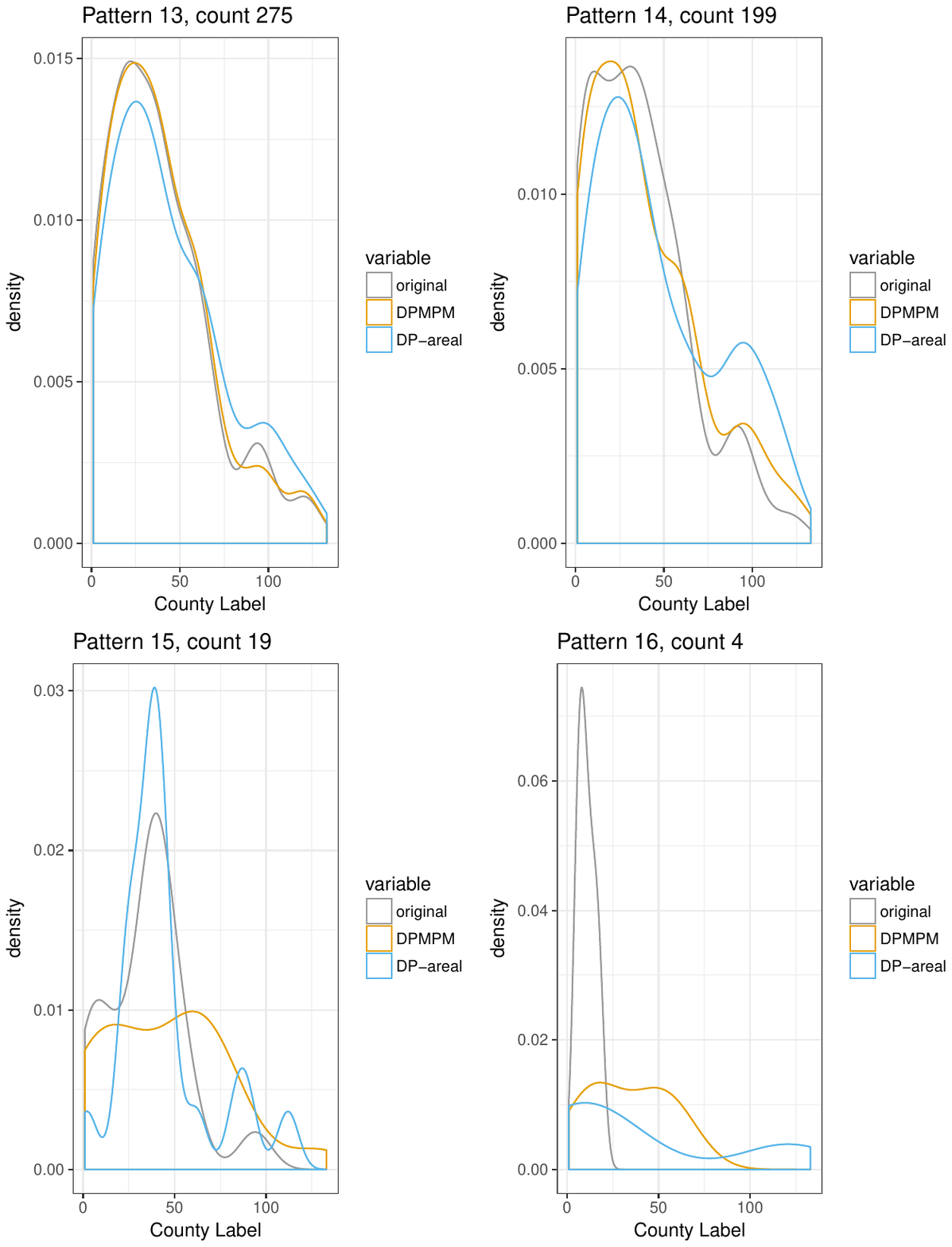}
\caption{Counties in Pattern 13 to Pattern 16.}
\label{fig:CountyLabelsPattern13to16}
\end{figure}

\begin{figure}[H]
\centering
\includegraphics[width=12cm, height=12cm]{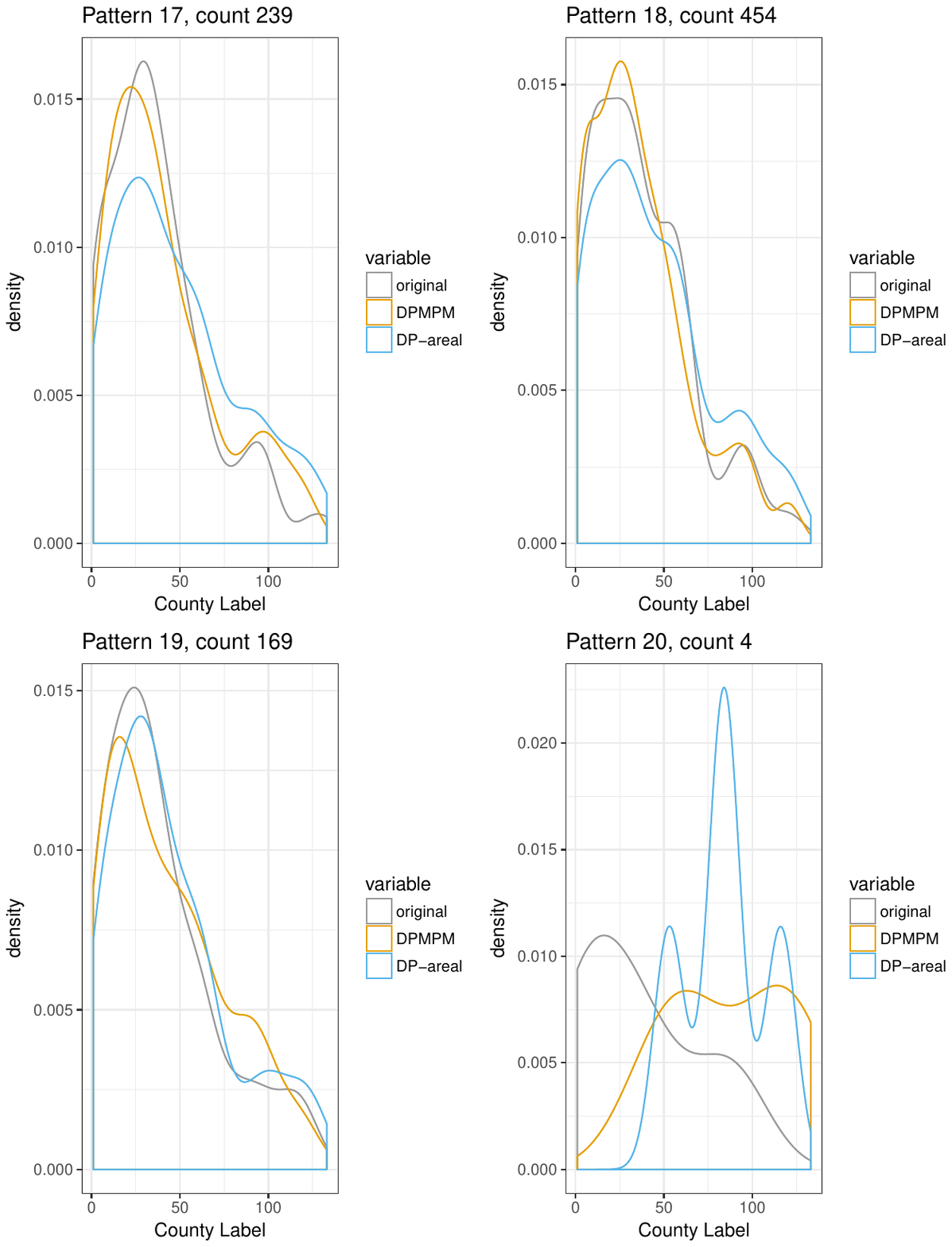}
\caption{Counties in Pattern 17 to Pattern 20.}
\label{fig:CountyLabelsPattern17to20}
\end{figure}

\begin{figure}[H]
\centering
\includegraphics[width=12cm, height=12cm]{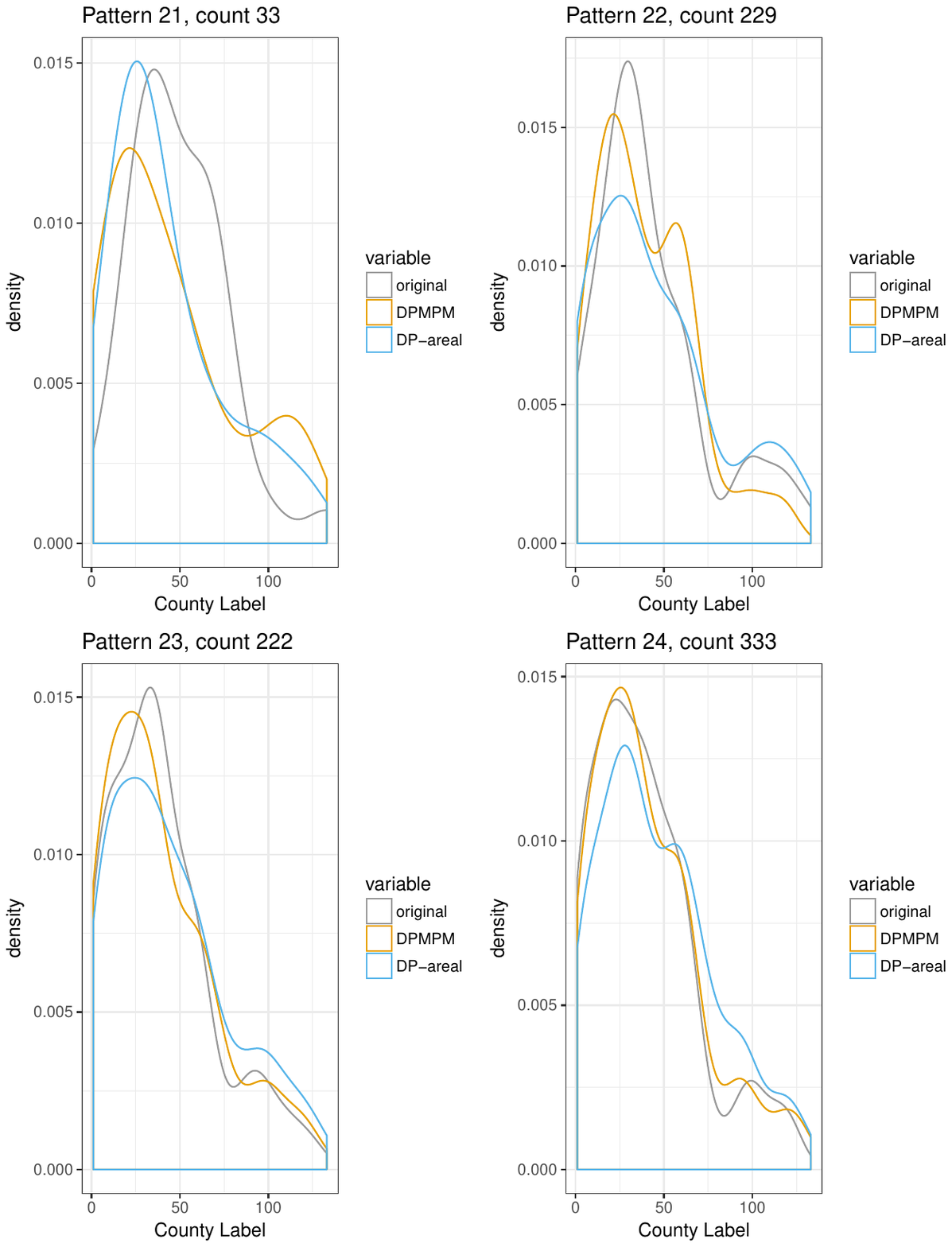}
\caption{Counties in Pattern 21 to Pattern 24.}
\label{fig:CountyLabelsPattern21to24}
\end{figure}

\begin{figure}[H]
\centering
\includegraphics[width=12cm, height=12cm]{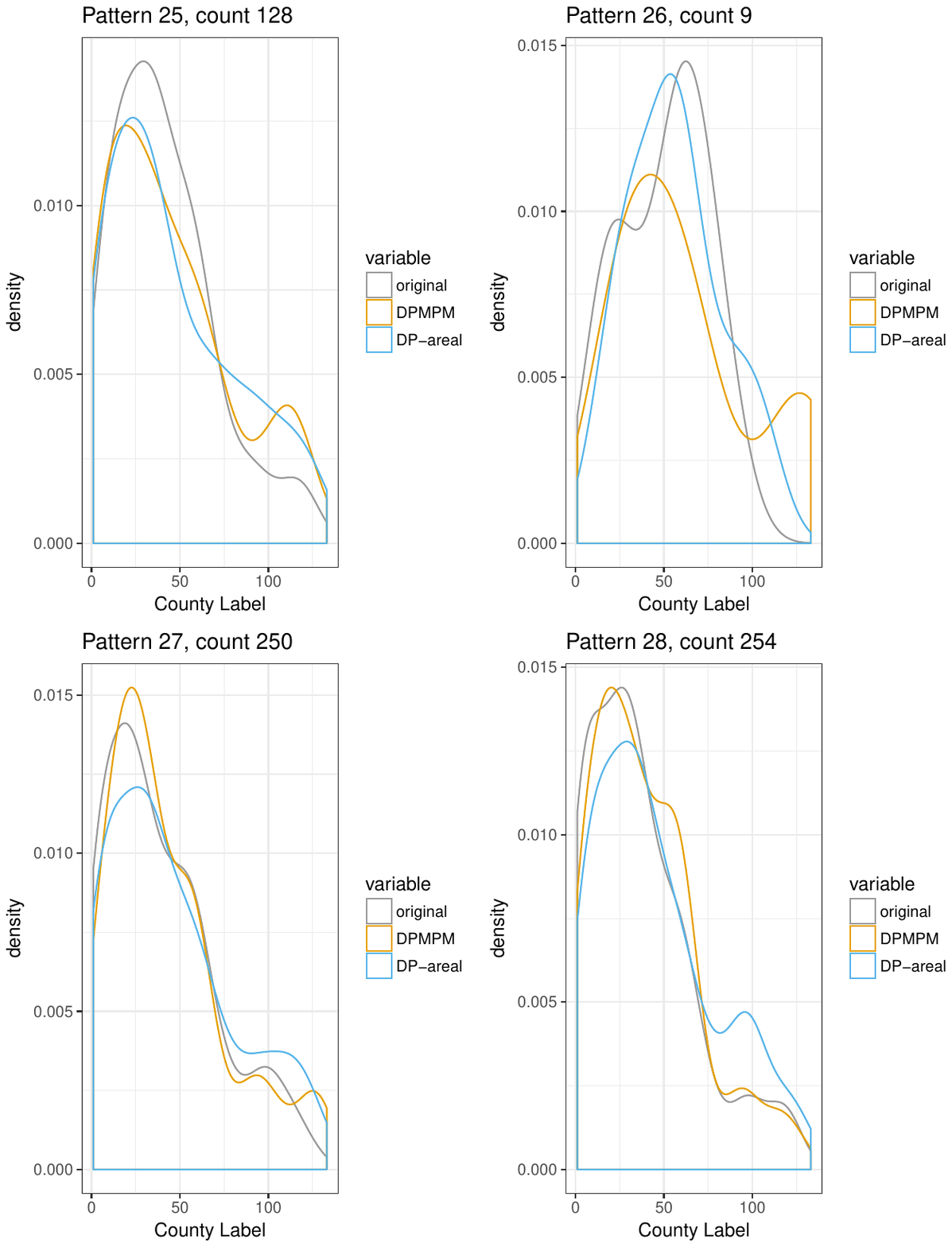}
\caption{Counties in Pattern 25 to Pattern 28.}
\label{fig:CountyLabelsPattern25to28}
\end{figure}

\begin{figure}[H]
\centering
\includegraphics[width=12cm, height=12cm]{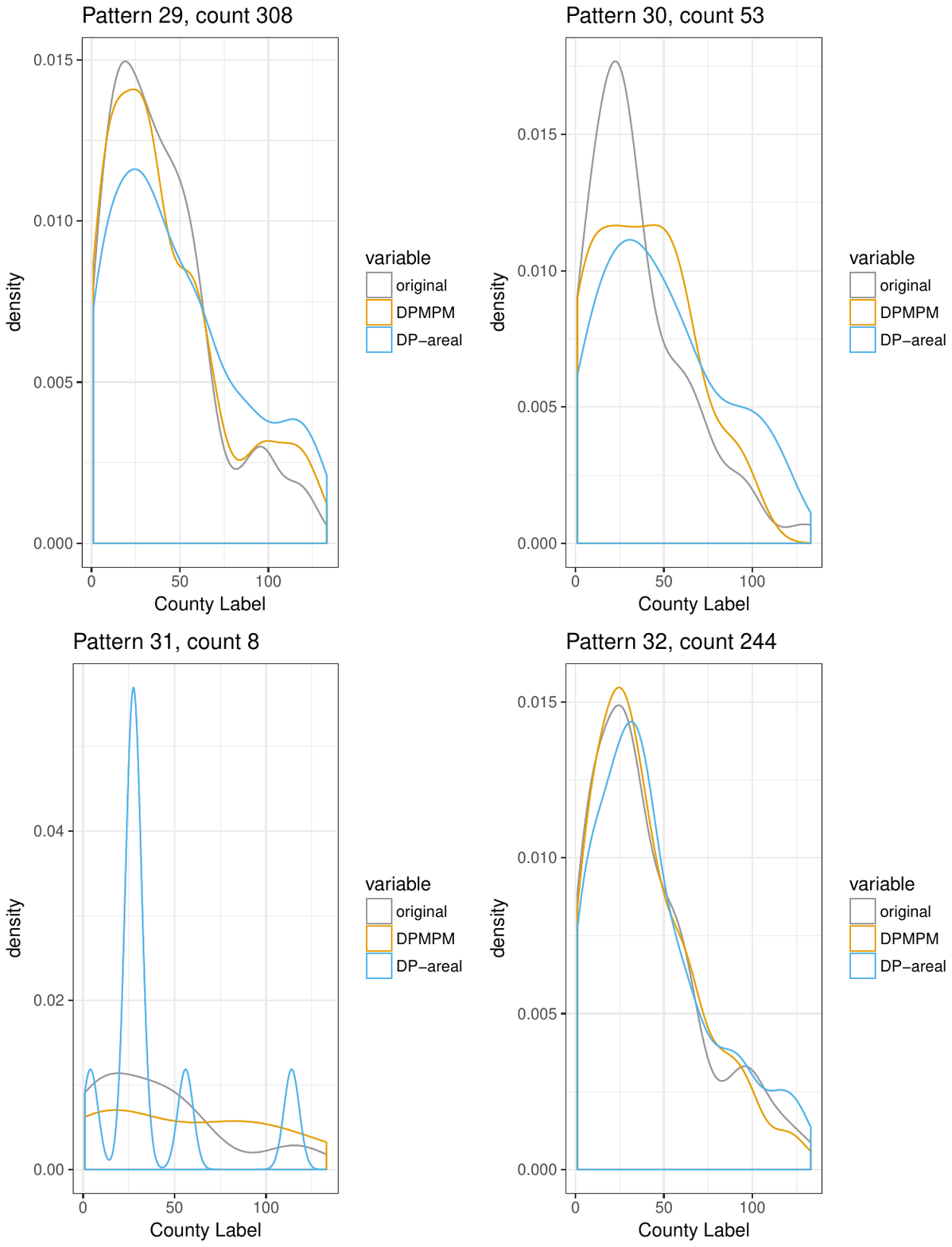}
\caption{Counties in Pattern 29 to Pattern 32.}
\label{fig:CountyLabelsPattern29to32}
\end{figure}

\begin{figure}[H]
\centering
\includegraphics[width=12cm, height=12cm]{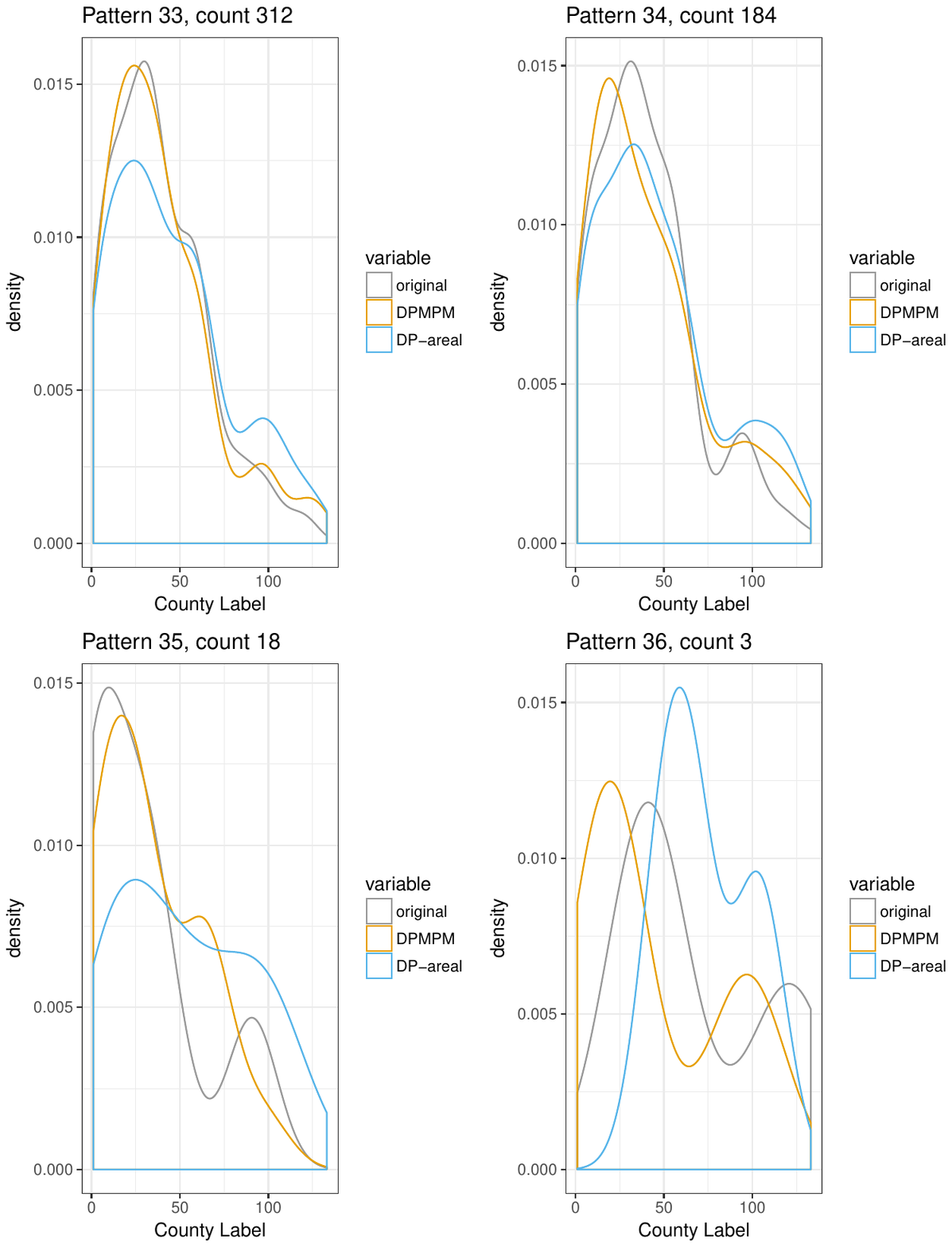}
\caption{Counties in Pattern 33 to Pattern 36.}
\label{fig:CountyLabelsPattern33to36}
\end{figure}

\begin{figure}[H]
\centering
\includegraphics[width=12cm, height=12cm]{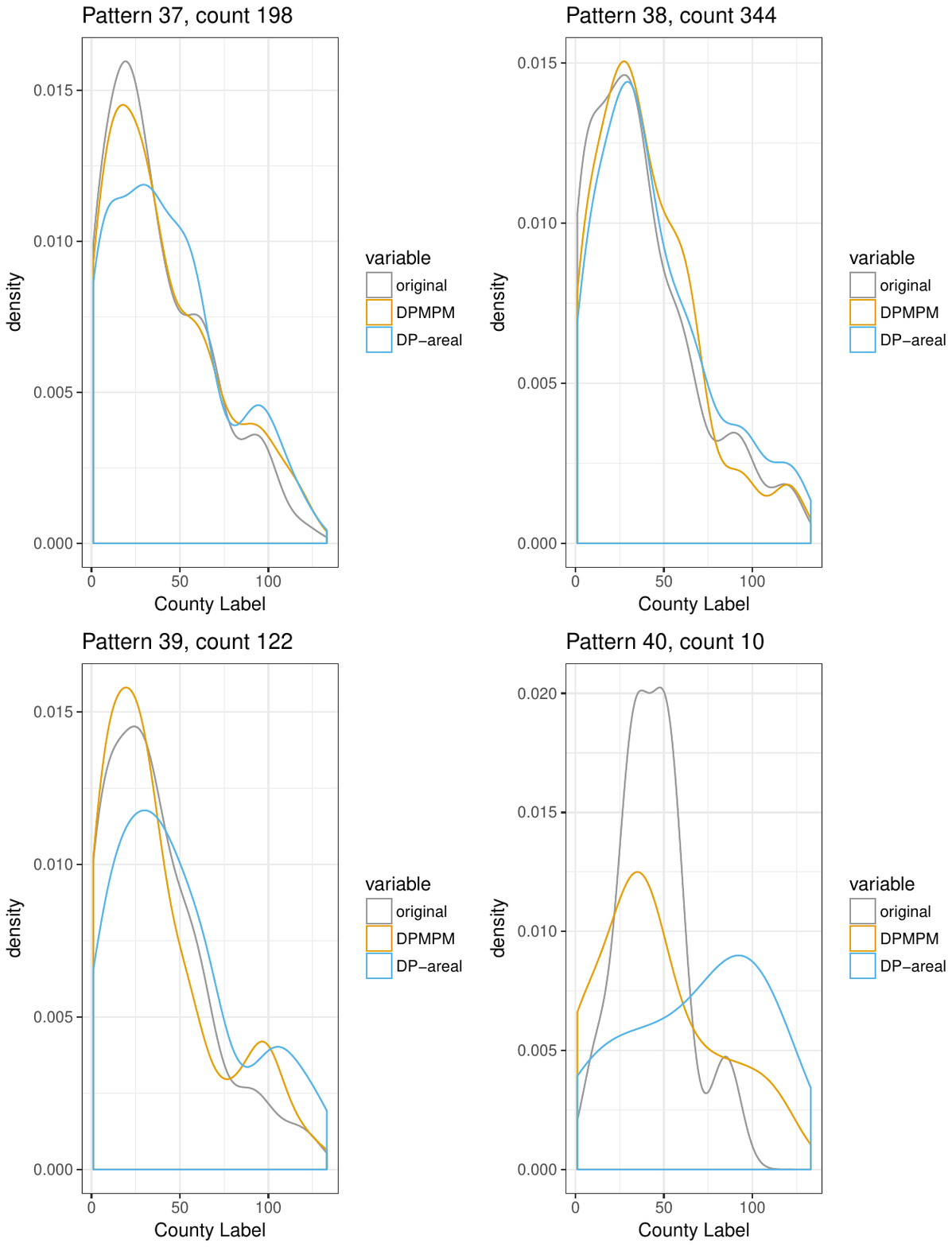}
\caption{Counties in Pattern 37 to Pattern 40.}
\label{fig:CountyLabelsPattern37to40}
\end{figure}

Figure \ref{fig:CountyLabelsPattern5to8} to Figure \ref{fig:CountyLabelsPattern37to40} are within pattern distribution plots of the county label synthesized from the DPMPM, DP-areal synthesizers and the original data distribution, from Pattern 5 to Pattern 40.

\newpage
\section{Identification Disclosure Risk Comparisons under Partially Observed Patterns}

\begin{figure}[H]
\centering
\includegraphics[width=8cm, height=10cm]{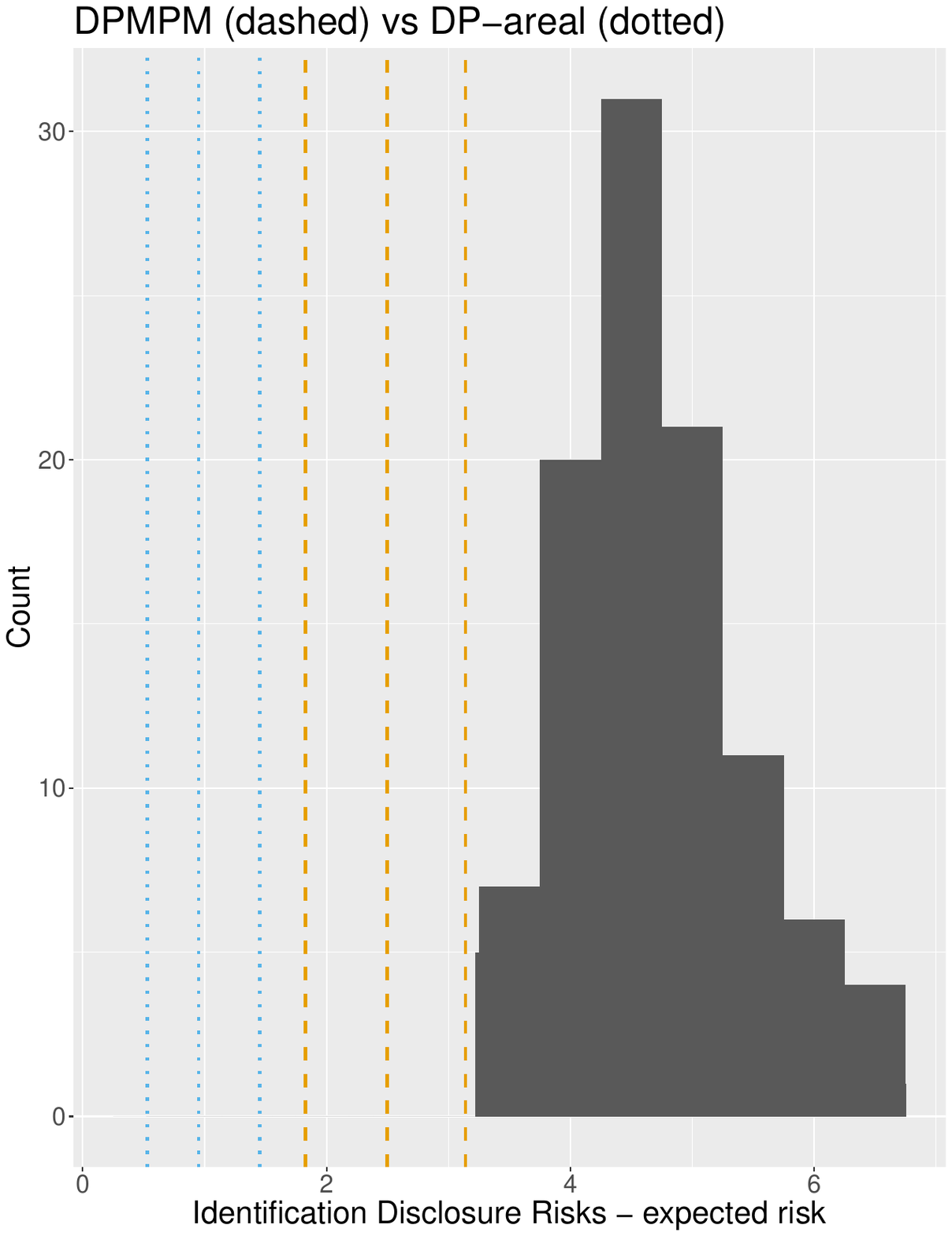}
\caption{Histogram of expected risks under the maximum risk scenario. Vertical lines include the min, mean, and max among the $m = 20$ synthetic datasets; dashed and orange for DPMPM, and dotted and blue for DP-areal. Known variables: gender and county label.}
\label{fig:IdMaxHist1}
\end{figure}

Figure \ref{fig:IdMaxHist1} is the set of of identification disclosure risks for the DPMPM, DP-areal and Maximum (from the original data) for the case where only gender and county label are known by the intruder.

\newpage
\section{Stan Script to Implement the DP-areal Synthesizer}

The following \texttt{Stan} script implements the DP-areal synthesizer:

{\footnotesize{
\begin{verbatim}
functions{
  
  real normalmix_lpdf(vector log_lambda, vector pi_prob, real mu, vector theta, 
  	matrix phi,  matrix X, real tau_lambda, int N, int K){
    real log_post;
    log_post = 0;
    for( i in 1:N ) /* by row of N x (R+1) */
    {
      vector[K] ps;
      for( k in 1:K )
      {
        ps[k]  = log(pi_prob[k]) + normal_lpdf(log_lambda[i]| mu + theta[k] + 
                                                 dot_product(phi[k],X[i]), 
                                                 	inv(sqrt(tau_lambda))); 
        
      } /* end loop k over clusters / mixture components */
        log_post += log_sum_exp(ps);
    } /* end loop i over N observations */
      
      return log_post;
  } /* end function normalmix_lpdf() */
    
} /* end function{} block */
  
  
  data{
    int<lower=1> N; // number of unique combinations of all attributes
    int<lower=1> K; // number of clusters
    int<lower=1> p; // number of non-geographic attributes
    row_vector[p] dj; // vector storing the number of levels of p non-geographic attributes
    int<lower=1> R; // number of total attribute levels: sum(dk)
    matrix[N,R] X; // each row is an R-by-1 vector comprising a one at position x_{ir}^{(b)}
    int c[N]; // set of N observations: the counts
  } /* end data block */
  
  
  transformed data{
    vector<lower=0>[K] ones_K;
    ones_K =  rep_vector(1,K); /* dirichlet prior on alpha has equal shapes */
  } /* end transformed parameters block */
  
  
  parameters{
    vector[N] log_lambda; /* poisson rates */
    real mu; /* global intercept */
    real alpha; /* DP concentration parameter on mixture model for point estimate */
    matrix[K,R] phi;
    vector[K] theta;
    simplex[K] pi_prob; /* mixture probabilities */
    vector[R] mu_phi;
    real<lower=0> tau_theta;
    real<lower=0> tau_phi;
    vector<lower=0>[R] sigma_phi;
    cholesky_factor_corr[R] L_phi;
    real<lower=0> tau_mu; /* precision in prior for mu */
    real<lower=0> tau_lambda; /* precision in prior for log_lambda[i] */
        
  } /* end parameters block */
  
  
  transformed parameters{
    vector[N] lambda; /* fitted values */
      
      for( i in 1:N )
      {
        lambda[i] = exp(log_lambda[i]);
      } /* end loop i over domains */
      
      
  } /* end transformed parameters block */
  
  model{
    
    // priors for cluster locations
    alpha           ~ gamma( 1.0, 1.0 ); /* DP concentration parameter */
    pi_prob         ~ dirichlet( alpha/K * ones_K ); /* instantiate a truncated DP prior */
      
      // normal prior for K x 1, theta
    theta           ~ normal(0,inv(sqrt(tau_theta))); /* vectorized */
    tau_theta       ~ gamma( 1.0, 1.0 );
    
    // multivariate Gaussian prior for R x 1, phi[k,]
    mu_phi          ~ normal(0,inv(sqrt(tau_phi))); /* vectorized */
    tau_phi         ~ gamma( 1.0, 1.0 );
    L_phi           ~ lkj_corr_cholesky(4);
    sigma_phi       ~ student_t(3,0,1); /* vectorized */
      for(k in 1:K )
      {
        /* phi[k] is the kth row of K x R, phi */
          to_vector(phi[k]) ~ multi_normal_cholesky(mu_phi, 
                                                     diag_pre_multiply(sigma_phi,L_phi));
      }
    
    mu              ~ normal(0,inv(sqrt(tau_mu)));
    tau_mu          ~ gamma( 1.0, 1.0 );
    
    // latent response (mean) likelihood on the log scale - mixture of normals prior
    // note that the normal prior allows for over-dispersion
    log_lambda      ~ normalmix(pi_prob, mu, theta, phi, X, tau_lambda, N, K);
    tau_lambda      ~ gamma( 1.0, 1.0 );
    
    // observed response likelihood
    c  ~ poisson_log(log_lambda);
    
    
  } /* end model{} block */
  
  
\end{verbatim}

}}
